\documentclass[fleqn,usenatbib]{mnras}

\usepackage{newtxtext,newtxmath}

\usepackage[T1]{fontenc}

\DeclareRobustCommand{\VAN}[3]{#2}
\let\VANthebibliography\thebibliography
\def\thebibliography{\DeclareRobustCommand{\VAN}[3]{##3}\VANthebibliography}

\usepackage{graphicx}	
\usepackage{amsmath}	

\usepackage{orcidlink}

\newcommand{\gizmo}{\textsc{gizmo}}
\newcommand{\slug}{\textsc{slug}}

\newcommand{\edit}{\textcolor{black}}
\newcommand{\newedit}{\textcolor{black}}

\newcommand{\aref}[1]{\hyperref[#1]{Appendix~\ref{#1}}}

\title[Action space coherence in the Galactic disc]{Evolution of action-space coherence in a Milky Way-like simulation}

\author[A. Arunima et al.]{
Arunima Arunima\orcidlink{0009-0009-2580-1027}$^{1}$\thanks{E-mail: arunima.arunima@anu.edu.au},
Mark R. Krumholz\orcidlink{0000-0003-3893-854X}$^{1},$
Michael J. Ireland$^{1},$
Chuhan Zhang$^{1}$
and Sven Buder$^{1}$
\\
$^{1}$Research School of Astronomy and Astrophysics, Australian National University, Canberra ACT 2601, Australia \\
}

\date{Accepted XXX. Received YYY; in original form ZZZ}

\pubyear{\the\year{}}

\begin{document}
\label{firstpage}
\pagerange{\pageref{firstpage}--\pageref{lastpage}}
\maketitle

\begin{abstract}
Efforts to dynamically trace stars back to the now-dissolved clusters in which they formed rely implicitly on the assumption that stellar orbital actions are conserved. While this holds in a static, axisymmetric potential, it is unknown how strongly the time-varying, non-axisymmetric structure of a real galactic disc drives action drift that inhibits cluster reconstruction.
We answer this question using a high-resolution magnetohydrodynamic simulation of a Milky Way-like spiral disc galaxy. We show that, while stars experience significant action evolution over $\lesssim 100$ Myr, they do so in a correlated fashion whereby stars born in close proximity maintain very similar actions for up to 0.5 Gyr. The degree of coherence shows no significant dependence on galactocentric radius, but varies between action components: vertical actions decohere for stars born more than a few hundred parsecs apart (likely due to giant molecular clouds), while radial and azimuthal actions remain correlated on kiloparsec scales (likely influenced by spiral arms). 
We use our measurements of the rate of action decoherence to develop a probabilistic framework that lets us infer the initial sizes of the star cluster progenitors of present-day \edit{`disc streams'} from their measured action distributions, which we apply to 438 known moving groups. Our results suggest that most of these streams likely originated from compact clusters, but that a significant minority are instead likely to be resonant or dynamically induced structures. This method of classifying streams complements existing methods, optimises the use of expensive spectroscopic abundance measurements, and will be enhanced by the more precise kinematic data that will soon become available from \textit{Gaia} DR4.
\end{abstract}



\defcitealias{arunima2025}{Paper I}

\section{Introduction}

Stars form in clustered environments inside giant molecular clouds (GMCs), inheriting mutually similar kinematics and chemical compositions from their shared birth material \citep{lada_lada2003}. Most of this clustered structure is not gravitationally bound following dispersal of the star-forming gas, and dissolves within tens to hundreds of million years (Myr) leaving the stellar population dispersed in the Galactic field. Even after the stars disperse in physical space, however, they retain imprints of their birthplaces in their characteristics \citep{freeman_bland-hawthorn2002}. If the dispersal occurs in a smooth, axisymmetric, and time-static potential, orbital actions are among these imprints,  as they are expected to remain adiabatic invariants under motion in a potential satisfying these conditions \citep{binney2008}. These conserved quantities have, hence, become central to dynamical studies and Galactic archaeology: if stars born together remain clustered in action space, then action-based co-natal tagging can become a powerful tool for mapping the star formation history of the Galaxy \citep{helmi2018, myeong2019, zucker22,swiggum25}.

However, the Milky Way potential is neither static nor axisymmetric. It is shaped by both external perturbations such as satellite interactions \citep{antoja2018,antoja2023,frankel2023} and secular internal processes such as transient spiral arm structure, the rotating Galactic bar, and the formation and dispersal of GMCs \citep{sellwoodbinney2002,roskar2008,fragkoudi2019,mackereth2019,tremaine2023}. These processes all contribute to a time-variable and non-axisymmetric component of the potential that can perturb stellar actions. Numerical simulations have confirmed this non-conservation of actions across a range of models: from $N-$body simulations without gas dynamics or star formation, the least realistic representations of the Galaxy \citep{solway2012,vera-ciro2014,mikkola2020}, to semi-analytic dynamical models \citep{kamdar2021}, and most recently and realistically, to fully self-consistent magnetohydrodynamical (MHD) simulations that include star formation, feedback and live disc with flocculent spiral arms \citep[][hereafter \citetalias{arunima2025}]{Tepper-Garcia22a, Tepper-Garcia25a, debattista2025,arunima2025}.

In \citetalias{arunima2025}, we quantified the non-conservation of individual stellar actions and showed that actions undergo a logarithmic random walk over timescales of hundreds of Myr. Yet a fundamental question still remains unanswered: Do co-natal stars retain coherence in action space over longer timescales? That is, while individual stars' actions undergo random walks that significantly change their values over timescales of $\sim 100$ Myr, is it possible that the random walks taken by stars that are born close together are \textit{correlated}, so that their actions remain close together in action space for longer times? There are compelling reasons to expect they might be. Co-natal stars, having shared initial conditions and experiencing same gravitational perturbations, should show similar changes in their actions and remain clustered in action space. If such coherence persists, it  would lend strong support to the use of actions for co-natal tagging, providing a robust tool for identifying disrupted clusters.

Indeed, several earlier studies have found evidence of long-lived clustering in phase space or integral-of-motion space even after spatial phase mixing. $N-$body studies of disrupting satellite galaxies or clusters have shown that stars can remain tightly grouped in action space for many orbital periods \citep{helmi2000,gomez2010}. Using cosmological hydrodynamical simulations, \citet{arora2022} demonstrated that stars born close together in the FIRE-2 Milky Way analogue retain action-space coherence for several hundred Myr before gradual dispersion in an evolving potential. Similarly, \citet{kamdar2019} modelled clustered star formation with realistic Galactic potential and found that co-natal stars maintain correlated kinematic signatures for extended timescales. Observationally, recent studies using \textit{Gaia} have identified groups with dynamical coherence indicative of common origin due to their chemical and age similarities \citep{coronado2020,furnkranz2024}. These results suggest that actions indeed hold promise as diagnostics of star-formation origins, but their utility depends on timescales, birth separation, and the dynamical environment.
Despite these advances, no study has yet examined the evolution of co-natal action coherence in a high-resolution, self-consistent MHD simulation of a live Milky Way-like disc. This gap leaves open uncertainties in how reliably actions can be used as archaeological tracers of stellar birth origins.

In this paper, we address this question using a high-resolution MHD simulation of a Milky Way-like disc galaxy. Building directly on our previous study of individual stars' action diffusion, we aim to extend the study of evolution of stellar actions for pairs of stars. This allows us to quantify the evolution of action differences and to measure the characteristic timescales and physical scales over which stars lose dynamical coherence in action space (a process we refer to as action-space decoherence). Through this, we can infer the co-natal origin of stars based solely on their current dynamical separation -- which is easier than ever to achieve thanks to immaculate astrometric information from \textit{Gaia} DR3 \citep{gaiadr3}, with the promise of even better data from \textit{Gaia} DR4 in the very near future.  Hence, we explore the use of our insights for observational analysis of star cluster disruption, stellar \edit{disc} streams\footnote{\edit{Usually in literature, the term \textit{stellar streams} is used to refer to structures associated with tidally disrupted globular clusters and dwarf galaxies accreted in the Galactic halo. In this work, however, we focus exclusively on kinematic structures present in the Galactic disc associated with open clusters or moving groups. We therefore adopt the term disc streams, following the usage of \citet{ratzenboeck2025}. Throughout this paper, the term streams refers exclusively to such disc streams unless stated otherwise.}}, dynamical reconstruction of co-natal groups in the Galactic disc. 

The paper is organised as follows. In \autoref{sec:methods}, we describe the galactic simulation and methodology used in this study. In \autoref{sec:results}, we present the main results from this study including temporal evolution of action differences for pairs of stars, scales of decoherence as well as dependence of this evolution on the stars' birth environment. In \autoref{sec:obs}, we outline an observational application of our findings to astrometric data from \textit{Gaia} and spectroscopic surveys, which should allow for inferences of birth conditions of moving groups and stellar streams in the Milky Way disc. Finally, in \autoref{sec:conclusions}, we summarise our key results and discuss their implications.

\section{Methods}
\label{sec:methods}

\subsection{Simulation}
\label{subsec:simulation}
\defcitealias{wibking2023}{WK23}
\defcitealias{chuhan2024}{Z25}
\defcitealias{hu2023}{H23}

We use the same high-resolution MHD simulation of an isolated Milky Way-like disc galaxy\edit{, which does not form a central bar but develops flocculent spiral structure,} previously employed in \citetalias{arunima2025}. This simulation, originally introduced by \citet[hereafter \citetalias{chuhan2024}]{chuhan2024}, is an extension of the full galaxy zoom-in simulations described by \citet{wibking2023} and \citet{hu2023} (hereafter \citetalias{wibking2023} and \citetalias{hu2023}, respectively). Here, we provide a brief summary of the simulation setup and refer readers to \citetalias{arunima2025}, \citetalias{wibking2023}, \citetalias{hu2023} and \citetalias{chuhan2024} for full details. 

The simulation follows the evolution of a turbulent galactic disc under ideal MHD using the \gizmo~code \citep{hopkins2015,hopkins2016,hopkins_raives2016} with radiative cooling implemented via the \textsc{grackle} library \citep{smith2017}. Star formation occurs stochastically when the gas particle density $\rho_{\text{g}}$ exceeds a critical threshold density $\rho_{\text{crit}}$; for particles meeting this condition the probability of being converted to a star particle in a time step of size $\Delta t$ is
\begin{equation}
    P = 1 - \text{exp}(-\epsilon_{\text{ff}}\Delta t/t_{\text{ff}}),
\end{equation}
which corresponds to setting the star formation rate density to 
\begin{equation}
    \dot{\rho}_{\text{SFR}} = \epsilon_{\text{ff}} \frac{\rho_{\text{g}}}{t_{\text{ff}}}.
\end{equation}
Here $\epsilon_{\text{ff}}$ is the star formation efficiency, $\rho_{\text{g}}$ is the local gas density and 
\begin{equation}
\label{eq:ff_time}
    t_{\text{ff}} = \sqrt{\frac{3\pi}{32G\rho}}
\end{equation} 
is the local gas free-fall time. The critical threshold density $\rho_{\text{crit}}$ is chosen to ensure that the local gravitational collapse is resolved, that is, the Jeans mass is nearly equal to the simulation mass resolution. The simulation uses a fixed efficiency per free fall time $\epsilon_{\rm ff} = 0.01$, motivated from observations \citep{Krumholz_etal2019}, increasing to $\epsilon_\mathrm{ff} = 10^6$ for gas particles with $\rho_{\text{g}} \geq 100 \rho_{\text{crit}}$; the latter choice is purely for numerical convenience, as it forces immediate conversion to star particles above this density threshold, thereby avoiding the extremely short timesteps that would otherwise be required in very dense gas.

The simulation mass resolution is high enough that stellar particles represent stellar populations too small to sample the full initial mass function (IMF), so the simulation treats stellar feedback on a star-by-star basis. When a star particle forms the code samples individual stars stochastically from a \cite{chabrier2005} IMF using the stellar population synthesis code \slug~\citep{dasilva2012,krumholz2015}. Thereafter, \slug~computes the age-dependent internal properties of each star from the Padova stellar tracks \citep{bressan2012}, and the corresponding ionising luminosity using the ``starburst99'' spectral synthesis method \citep{leitherer1999}. The feedback module also uses \slug~to determine which stars end their lives as supernovae versus as AGB stars following the models of \citet{sukhbold2016}. The simulation deposits photoionisation and supernova feedback in the gas using the approaches described by \cite{Hopkins:2018} and \cite{armillotta2019}, and returns mass and metals following the predictions of \cite{sukhbold2016} for type II supernovae, \cite{karakas_lugaro2016} for AGBs, and \cite{doherty2014} for super-AGBs. 

\begin{figure}
    \includegraphics[width=\linewidth]{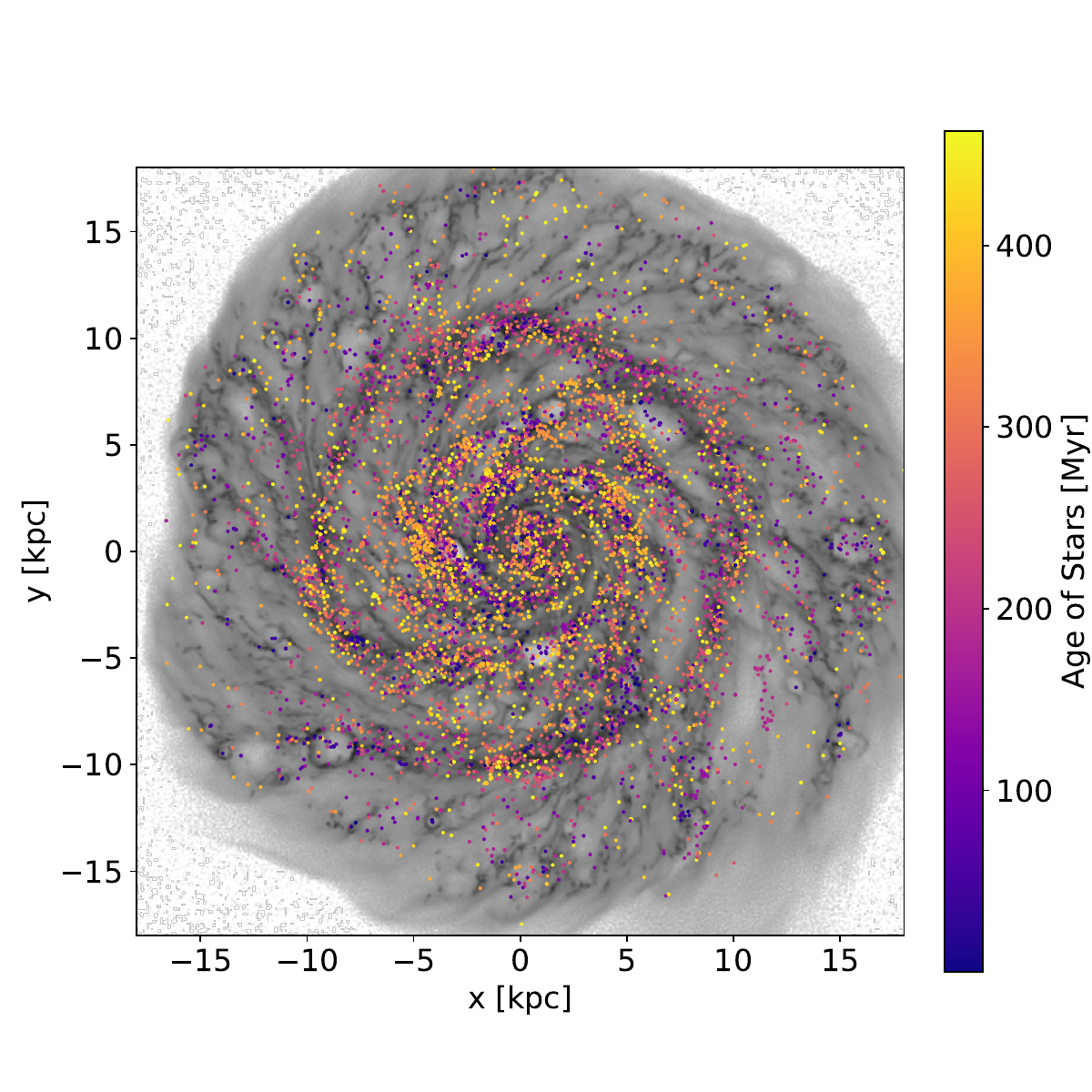}
    \caption{\edit{A snapshot of the simulation at $t=464$ Myr (the final snapshot). We show the log of the gas surface density in greyscale in the background, with darker shades corresponding to higher densities. Overlaid on this is a random subsample representing 1\% of the total star sample we use in this work, with the colour of each particle indicating its age as shown by the colour bar.}}
    \label{fig:snapshot}
    
\end{figure}

The simulation proceeds in two stages. The first phase, described in \citetalias{wibking2023}, evolves an isolated galaxy from AGORA Milky Way-like initial conditions \citep{Kim:2016} using IMF-integrated feedback. \edit{The \citetalias{wibking2023} simulation includes a live dark matter halo, a stellar disc and bulge, and a gas disc. Gas particles have a mass of 859.3 M$_\odot$ contributing to a gas disc of mass $3.4 \times 10^{10}$ M$_\odot$, while dark matter particles have a mass of $1.254 \times 10^5$ M$_\odot$, contributing to a halo of mass $1.07 \times 10^{12}$ M$_\odot$. Stellar disc and bulge particles have a mass of $3.4373 \times 10^3$ M$_\odot$, contributing to a stellar disc of mass $3.4 \times 10^{10}$ M$_\odot$ and a bulge of mass $4.3 \times 10^9$ M$_\odot$. The stellar disc has a scale height of approximately 350 pc, while the gas disc follows a double-exponential profile with a radial scale length of 3.43 kpc and a vertical scale height of 343.2 pc.} 

The second phase is a high resolution re-simulation: using the 600 Myr snapshot from the \citetalias{wibking2023} simulation (which exhibits a stable gas fraction similar to the present day Milky Way), \citetalias{chuhan2024} run the simulation for 100 Myr without enhancing the resolution but switching from the IMF-integrated treatment of feedback to the star-by-star feedback prescription described above. Then, the \edit{gas} mass resolution is increased to $\Delta M \approx 300 {\rm M}_\odot$ using the particle splitting procedure outlined in \citetalias{hu2023} and the simulation continues for another 464 Myr, with snapshots output at 1 Myr intervals. By the end of the simulation, approximately 1.32 million star particles have formed.

\edit{In \autoref{fig:snapshot}, we show the state of the simulation in the final snapshot. The greyscale background shows the gas surface density, with darker regions indicating higher densities, while the coloured points show 1\% of the 1.32 million star particles, selected at random and coloured according to their ages as shown by the accompanying colour bar. The disc clearly has a self-consistently-generated multi-armed flocculent spiral structure rather than a long-lived grand-design pattern. At the resolution of the re-simulation phase, star formation occurs within self-gravitating GMCs that form self-consistently from the turbulent interstellar medium. The resulting GMC population and its disruption by stellar feedback are consistent with the Galactic tuning-fork test \citep{zipeng2024}, which characterises the scale-dependent spatial decorrelation between molecular gas and H$\alpha$ emission \citep{kim2022}. This demonstrates that the GMC properties in our simulation are physically reasonable, a point that will be important to keep in mind below when we discuss how the gravitational perturbations induced by these GMCs influence stellar orbits.}

While \edit{the ``star particles'' in our simulation} represent small sub-cluster-scale populations rather than true individual stars, their masses are low enough that they should preserve the dynamical information imprinted at birth almost as well as individual stars. At the smallest scales, our simulation is expected to overestimate decoherence in action space due to star particles being significantly larger than median stellar masses.

\subsection{Action calculation}
\label{subsec: action calc}
Our first step in analysing these simulations is to compute actions for each star particle at each output time. Our full procedure for doing so, including several tests of the method, is provided in Section 3 and the Appendices of \citetalias{arunima2025}. We therefore only summarise here for reader convenience.

We calculate actions for the star particles using the epicyclic approximation, which is very accurate for young stars near the disc plane and is much more computationally efficient than alternative approaches. At each simulation snapshot, we use the gravitational potential computed by \textsc{gizmo} at the position of each particle to construct an axisymmetric model for the galaxy's potential $\Phi$. Following \cite{binney2008}, we then decompose the stellar orbits into independent radial and vertical oscillations about a guiding centre at galactocentric radius $R_g$, characterised by the radial ($\kappa$) and vertical ($\nu$) frequencies
\begin{eqnarray}
    \kappa^2 & = & \left( R \frac{\partial \Omega^2}{\partial R} + 4\Omega^2\right)_{(R=R_g, z=0)} 
    \label{eq:kappa} \\
    \nu^2 & = & \left. \frac{\partial^2 \Phi}{\partial z^2}\right|_{(R=R_g, z=0)},
    \label{eq:nu}
\end{eqnarray}
where 
\begin{equation}
    \Omega (R) = \sqrt{\frac{1}{R} \left. \frac{\partial \Phi}{\partial R} \right|_{R, z=0}}
    \label{eq:Omega}
\end{equation}
is the circular frequency.

To assign actions to each star, we first evaluate the stellar guiding centre from the star's angular momentum $L_z$ by solving the implicit equation
\begin{equation}
 |L_z| = R|v_{\phi}| = R_g^2 \Omega (R_g)
 \label{eq:Rg}
\end{equation}
for $R_g$, where $R$ is the star's radial position and $v_{\phi}$ its azimuthal velocity. From $R_g$ we can immediately evaluate $\kappa$ and $\nu$ from \autoref{eq:kappa} and \autoref{eq:nu}. We then compute the stellar radial, vertical, and azimuthal actions as
\begin{eqnarray}
    J_R &= & \frac{v_R^2 + \kappa^2 ( R - R_g)^2}{2\kappa}
    \label{eq:Jr} \\
    J_z &=& \frac{v_z^2 + \nu^2z^2}{2\nu}
    \label{eq:Jz}\\
    J_{\phi} &=& L_z.
    \label{eq:Jphi}
\end{eqnarray}

Finally, as mentioned in \citetalias{arunima2025}, many stars in the simulation are part of gravitationally bound clusters for at least some of their evolution, and the internal orbital motion within these clusters drives large variations in galactocentric actions. These can obscure the slower evolution induced by the Galactic potential, which is of primary concern here. To remove these short-timescale fluctuations, we apply a Butterworth low-pass filter with a cut-off frequency of 1/30 Myr$^{-1}$ to the action time series of each star. We use these filtered actions for all the analysis in this paper.

\subsection{\edit{Selecting the stellar sample}}
\label{subsec:NNsec}

\begin{figure}
    \includegraphics[width=\linewidth]{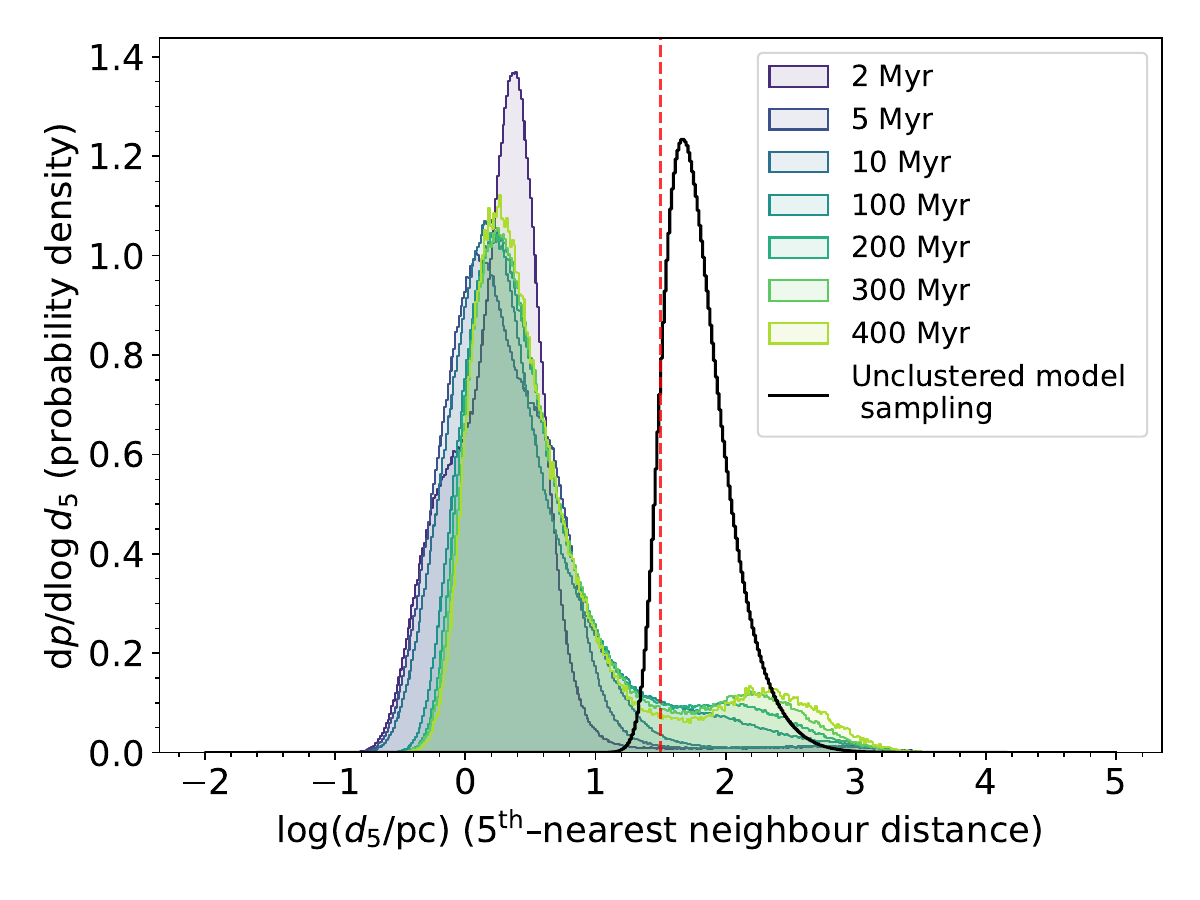}
    \caption{The distribution of 5th nearest neighbour distances $d_5$ for coeval stars at different ages (shown in the legend). \edit{For comparison we also show as a black solid line the distribution of $d_5$ for an un-clustered stellar population sampled from a double exponential disc model (see main text), with its normalisation rescaled for easier visual comparison.} We show the 30 pc limit that we use to separate `clustered' stars from dispersed stars as the red dashed vertical line.}
    \label{fig:NNdist}
    
\end{figure}

The overarching goal of this work is to determine whether stars born together remain clustered in action space even as they drift apart in physical space. To this end, at each snapshot, we begin by \edit{quantifying stellar distances at birth, so that we can find ones that formed nearby to one another.}
We first identify stars that were born within the last 1 Myr and consider all pairs of two such stars. We group our pairs based on the physical distance between the two stars in the first snapshot in which the stars appear, $d_\text{init}$. We are interested in knowing how the distance in \textit{action} between each pair of stars grows with the time $\Delta t$ since their birth, as a function of their \textit{physical} separation at birth $d_\text{init}$.

\edit{The next step is to identify stars that have separated in physical space as they age, and particularly to filter out any stars that have remained together for long times because they are part of gravitationally-bound clusters -- these will remain clustered in action space simply because of their boundedness.}
To define a criterion by which we can identify dispersed stars\edit{, we start by examining} the distribution of 5th nearest-neighbour distances for coeval stars -- again defined as stars born in the same 1 Myr cohort -- measured at various stellar ages. 
\edit{(We verify in \aref{app:NN} that choosing a different neighbour number rather than 5th nearest yields qualitatively identical results.) The age cohorts are constructed across all snapshots of the simulation -- thus, for example, the distribution for the 2 Myr cohort plotted in \autoref{fig:NNdist} is measured from the stars with ages of 2-3 Myr present in all the snapshots we analyse, rather than those selected from a single snapshot at a fixed time. As a result, each individual star particle contributes to multiple age cohorts over the course of the simulation, and each cohort contains $\sim 10^6$ star particles, ensuring excellent sampling statistics.}

\edit{In order to interpret the results shown in \autoref{fig:NNdist}, it is helpful to construct a comparison example showing the expected separation distribution for stars whose overall distribution matches that in the simulation, but with no clustering. To construct this comparison sample, we fit the distribution of star particle position in the simulation (including only particles formed self-consistently in the simulation, excluding those present at $t=0$) with a double-exponential model with surface density $\Sigma(R) \propto \exp(-R/R_d)$ and vertical density $\rho(z) \propto \exp(-|z|/H)$ where $R_d$ is the radial scale length and $H$ is the disc scale height. We carry out this fit for each simulation snapshot, but find that there is very little variation in $R_d$ or $H$ from one snapshot to another, as expected since the disc overall is in statistical steady state. We therefore adopt the median values $R_d =3.09$ kpc and $H = 153$ pc across all snapshots. We then sample $1.2 \times 10^6$ star particles from this distribution, matching the mean number present in each of our age cohorts, and compute their fifth-nearest neighbour distance distribution. We repeat this procedure 100 times, and plot the average result as the black line in \autoref{fig:NNdist}.}

\edit{We see that the control distribution peaks at separations of $\sim 50$ pc, which we interpret as the typical interstellar separation expected for stars that are not clustered but instead populate the galactic disc according to its large-scale density structure. By comparison, for all the age cohorts of newborn stars in our simulations,} the peak of the fifth-nearest neighbour distance is much smaller than this, $\sim 1$ pc, reflecting that stars are clustered together. At the earliest ages the distribution of separations is unimodal and consists only of this clustered feature, but at later times a secondary feature \newedit{that is near the peak of the control distribution} begins to appear, and the separation distribution becomes bimodal; the minimum between these two modes is at $\approx 30$ pc. We therefore identify the small-separation peak \edit{as arising from stars that remain close to the clusters or in compact associations} and the large-separation peak \edit{as reflecting stars that have dispersed from their natal environments into the galactic field, and are approaching a spatial distribution consistent with no clustering beyond that imposed by the large-scale stellar distribution in the disc.\footnote{\edit{Careful readers may notice that the mean separation of the second peak is slightly larger than that of the comparison sample. This difference arises because this second peak contains significantly fewer than the 1.2 million stars -- the size of the full cohort sample -- used to construct the comparison unclustered model. An unclustered sample containing $\approx 200,000$ stars, about the number in the second peak for the 400 Myr cohort, would have a mean separation that is larger by a factor of $6^{1/3}\approx 1.87$, closely matching what we observe.}} Based on this, we adopt a conservative criterion and} retain in our sample only those stellar pairs for which the separation between the stars is $>30$ pc at time $\Delta t = 100$ Myr after the stars' birth. \edit{This choice intentionally errs on the side of excluding potentially coherent but unbound structures and} effectively limits our sample to pairs of stars that are no longer recognisable as members of the same bound cluster in physical space. \edit{We further test the robustness of our results to this sample selection by adopting alternative separation thresholds in \aref{app:threshold} and find no practical differences in the inferred behaviour.}


\subsection{Quantifying action-space coherence}
\label{subsec:definitions}

We are now in a position to quantify the change in action differences between the pairs of stars that we retain. For a given pair of stars at each $\Delta t$, we define the absolute change in action difference for each component ($J_R$, $J_\phi$, $J_z$) as the change in their action difference at age $\Delta t$ from the action difference at the birth snapshot:
\begin{equation}
\delta_{\text{abs}}\Delta J (t,\Delta t) = \left|\left[\left|J_i (t+\Delta t) - J_j(t+\Delta t)\right| - \left|J_i (t) - J_j(t)\right|\right]\right|
    \label{eq:abs_action}
\end{equation}
where $J_{i} (t)$ is one of the components of action of the $i$th star at time $t$. Thus for example consider a pair of stars born at time $t$ whose radial actions at this time are $J_{R,i}(t) = 2$ km s$^{-1}$ kpc and $J_{R,j}(t) = 3$ km s$^{-1}$ kpc, and at a time $\Delta t = 100$ Myr later we find that their actions are $J_{R,i}(t+\Delta t) = 1$ km s$^{-1}$ kpc and $J_{R,j}(t+\Delta t) = 4$ km s$^{-1}$ kpc. We would then say that $\delta_\mathrm{abs}\Delta J_R(t, \Delta t) = 2$ km s$^{-1}$ kpc for this stellar pair, since the separation in actions has grown from 1 km s$^{-1}$ kpc at birth to 3 km s$^{-1}$ kpc at 100 Myr after birth. We also define relative change in action difference as the absolute change in action difference divided by the geometric mean of the actions of the stars at the later time:
\begin{equation}
    \label{eq:rel_action}
\delta_{\text{rel}}\Delta J(t,\Delta t) = \frac{\delta_{\text{abs}}\Delta J (t,\Delta t)}{\sqrt{J_i (t+\Delta t) J_j(t+\Delta t)}}
\end{equation}
For our example stellar pair, we would have \edit{$\delta_\mathrm{rel}\Delta J_R(t,\Delta t) = 1$}.


\section{Results}
\label{sec:results}
We are now prepared to characterise the evolution of action-space coherence for coeval stars. Our aim is to establish whether stars that are born together remain clustered in action space, and on what spatial and temporal scales this coherence persists. We begin by examining the overall behaviour of action differences of pairs as a function of their birth separation in \autoref{subsec:clustering}. In \autoref{subsec:decoherence}, we distinguish how coherence differs across the three action components. Finally, we test whether the rate of decoherence depends on the Galactic environment by repeating the analysis in radial bins across the disc in \autoref{subsec:radial}.

\subsection{Stars born together remain clustered in action space}
\label{subsec:clustering}

\begin{figure}
    \includegraphics[width=0.95\linewidth]{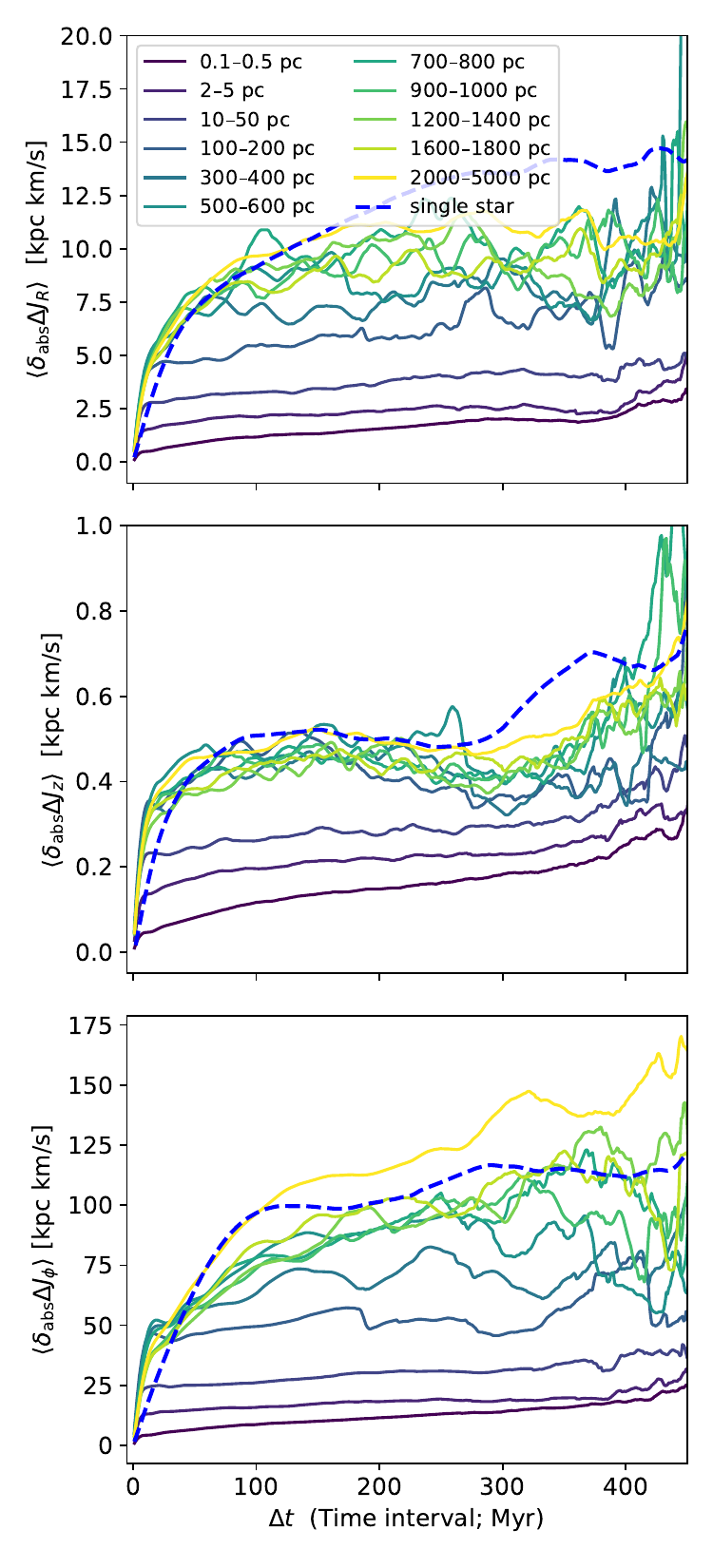}
    \caption{Median of absolute change in action difference (row-wise in order: $\delta_{\text{abs}}\Delta J_R, \delta_{\text{abs}}\Delta J_z$ and $\delta_{\text{abs}}\Delta J_{\phi}$) versus time for coeval pairs of stars binned by stellar separation at birth; birth distance bins are indicated by colour, as shown in the legend. For comparison, the blue dashed line shows the \edit{median absolute change in the action of \textit{single} stars from their own initial actions} as a function of age gap; this result is taken from from \citetalias{arunima2025}. As expected, for stars \edit{born} at large separations (lighter-coloured solid lines) the actions change in essentially uncorrelated ways, so the changes in action between pairs of stars are similar to the changes for individual stars (blue dashed line), but stars born close together (darker solid lines) change their actions in correlated ways, and so the \edit{median} change in action for pairs of stars is much smaller than the \edit{median} change in actions for single stars.}
    \label{fig:abs_change}
\end{figure}

\begin{figure}
    \includegraphics[width=0.95\linewidth]{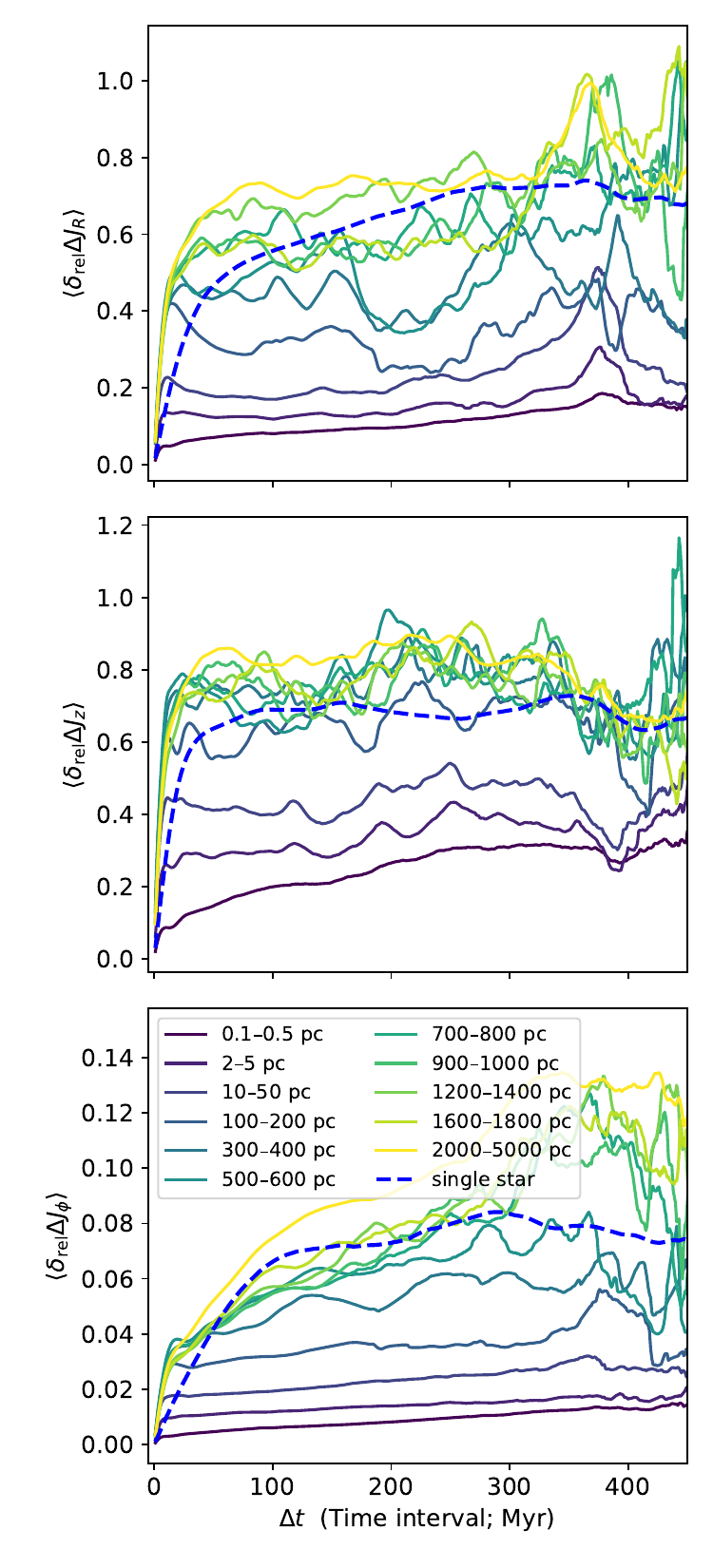}
    \caption{Same as \autoref{fig:abs_change}, but now showing changes in relative action $\delta_\mathrm{rel}\Delta J_{R,z,\phi}$ rather than absolute action $\delta_\mathrm{abs}\Delta J_{R,z,\phi}$.}
    \label{fig:rel_change}
\end{figure}

As described in the previous section, for each of our 1 Myr snapshots, we identify pairs of stars born since the previous snapshot, and record their physical separation in the first snapshot at which we find them, $d_\mathrm{init}$. We then track the actions of each star over time and calculate the absolute and relative changes in action differences for each stellar pair at each snapshot. We do this throughout the simulation, and the result is data set of stellar pairs at different times; each entry in this data set consists of the pair's distance at birth $d_\mathrm{init}$, the time $\Delta t$ since the pair was born, and the absolute and relative change in the pair's action difference since birth, $\delta_\mathrm{abs}\Delta J_{R,z,\phi}$ and $\delta_\mathrm{rel}\Delta J_{R,z,\phi}$. To study how these action changes depend on birth distance, we bin this data set by $d_\mathrm{init}$, and for each time interval $\Delta t$ in each $d_\mathrm{init}$ bin, we compute the median of the changes in action differences for all stellar pairs. We plot these median action differences for each component as a function of time in \autoref{fig:abs_change} for absolute action changes, and in \autoref{fig:rel_change} for relative changes. 

From the figures, it is clear that the stars born close together show much smaller change in action differences over time compared to the ones born far apart. That is, we find that at any given $\Delta t$, the absolute and relative changes of all actions are much smaller for the small $d_\mathrm{init}$ bins than for the larger ones. As the birth separation increases, the curves for different bins begin to converge (in both absolute and relative changes in the action differences), suggesting that the pairs of stars born sufficiently far apart evolve independently in action space, i.e. they are decoherent from birth.

To contextualise these changes, we also show the typical single-star action evolution from \citetalias{arunima2025} in \autoref{fig:abs_change} and \autoref{fig:rel_change} as the blue dashed lines. This quantity is defined as the change in the action of a single star relative to its initial action, so for example the blue dashed line in the top panel of \autoref{fig:abs_change} indicates that, for the median star in the simulations, its radial action $150$ Myr after birth will be $\approx 10$ kpc km/s different than it was at birth. The plot shows that the single-star action change with respect to itself in time closely matches the change in action difference for pairs of stars that are born far apart, confirming that these stars undergo uncorrelated evolution in action space. By contrast, pairs born close together retain coherence in action space over the entire $\sim 0.5$ Gyr timescale of our simulation. We emphasise that this does not mean that these stars' individual actions do not change -- as the blue dashed line shows, their individual actions are in fact changing significantly. It is simply that for stars that were born close together in space, the changes in action they experience are very similar, so the change in the difference of actions between the two stars is much less than the action change each star undergoes individually. We can conceptualise this as the stars undergoing a random walk but doing so while holding hands: after a long enough time the two stars may both wind up very far from where they started, but they will still be close to each other. For the stars in the smallest bins of $d_\mathrm{init}$, the relative action difference change remains well below 50\% throughout the simulation (\autoref{fig:rel_change}). This establishes that stars born together remain together in action space over at least hundreds of millions of years, even as they drift apart physically.

\subsection{Different components decohere on different spatial scales}
\label{subsec:decoherence}

A second quantity that we can read off from \autoref{fig:abs_change} and \autoref{fig:rel_change} is how closely together stellar pairs must be at birth for their actions to separate noticeably more slowly than the rates at which the actions of single stars drift -- that is, for what values of $d_\mathrm{init}$ do we observe that the change in action separation for stellar pairs (solid lines) is significantly smaller than the rates at which individual stellar actions drift (dashed blue lines)? We can think of stars with $d_\mathrm{init}$ smaller than this value as coherent in action space, while those with larger $d_\mathrm{init}$ are incoherent.
Interestingly, for radial and azimuthal actions, decoherence emerges gradually at birth separations on the scales of kiloparsecs. However, for vertical action, coherence is lost at much shorter distances of hundreds of parsecs. 

This difference is physically significant. In \citetalias{arunima2025}, we argued that vertical actions and radial actions have different drivers of change: vertical actions are primarily influenced by local perturbations such as GMCs, while radial actions are affected by large-scale structures such as spiral arms. This picture is consistent with the spatial scales of decoherence that we find here: the vertical action decoheres over scales comparable to the vertical disc scale height \citep{bobylev2021,vieira2023}, while radial and azimuthal actions decohere over much larger distances, reflecting the scale of spiral arm structure, which is comparable to the radial scale length of the stellar disc \citep{spiralarms?2021, bobylev2021}. \edit{Further support for this picture comes from the self-consistent, isolated disc simulation of \citet{debattista2025} in which all stars form from gas. They show that spiral structure strongly affects radial actions (derived under the assumption of axisymmetry) but leaves vertical actions comparatively unchanged.}

These results suggest that the coherence in action space is a function of birth distance but also depends on which action component is considered and what physical processes dominate its evolution. These dependencies \edit{provide} quantitative constraints on how long and how far action clustering persists in each component of action. As we explore in \autoref{sec:obs}, this can directly inform efforts to reconstruct disrupted stellar associations in the Milky Way by assessing the maximum allowable action difference as a function of birth separation and time since formation.

\subsection{Radial Dependence}
\label{subsec:radial}
To test whether action decoherence depends on Galactic environment, we repeat our analysis in radial bins of 1 kpc. In \citetalias{arunima2025}, we found that the relative action change was faster and stronger in the inner disc compared to the outer disc, even after removing the expected dependence due to radial variation of the galactic orbital period, as expected given the large number of perturbation drivers (e.g., higher density of GMCs) in the inner disc.

\begin{figure}
    \includegraphics[width=\columnwidth]{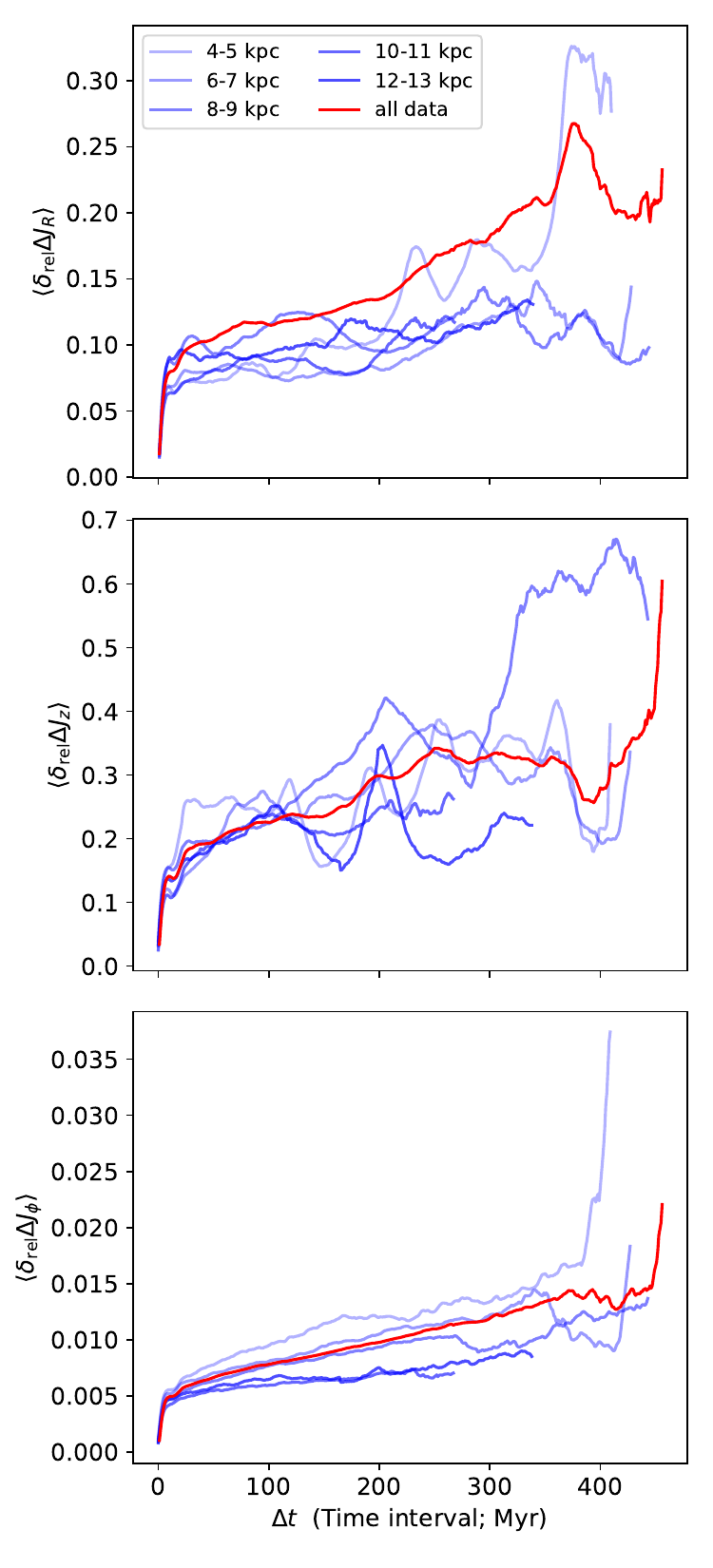}
    \caption{Median of relative change in action difference (row-wise in order: $\delta_{\text{rel}}\Delta J_R, \delta_{\text{rel}}\Delta J_z$ and $\delta_{\text{rel}}\Delta J_{\phi}$ in order) for coeval pairs of stars in time with their birth distances in the 0.5 -- 2 pc bin. The darker blues represent higher radii. \edit{We only plot the combinations of radial bins and time intervals that have more than $10^4$ star particles' pairs to show statistically significant trends.} The red curve shows the median of the relative change in action differences for the entire dataset of pairs of coeval stars born within 0.5 -- 2 pc of each other.}
    \label{fig:radial}
\end{figure}

Here, however, we restrict ourselves to stars that are born close together. In \autoref{subsec:clustering}, we saw that such stars remain clustered in action space throughout the simulation. For each radial bin, we select coeval stars with birth separations of 0.5 -- 2 pc and compute the median of relative change in action difference at each $\Delta t$; results for other small $d_\mathrm{init}$ bins are qualitatively similar, so we show only the 0.5 -- 2 pc bin for clarity. \edit{To ensure statistical robustness, we only plot radial bin and time combinations that contain more than $10^4$ stellar pairs.} The results are plotted in \autoref{fig:radial} for all three components of action, for radial bins of 1 kpc spanning 2 -- 19 kpc (the range where actions are available). \edit{We find little systematic radial dependence in the change of action difference across most of the disc. The most notable deviation occurs in the 4--5 kpc bin, which exhibits an increase in the median relative action change by up to a factor of $\sim 3$ at bigger time intervals compared to other radii. Although there is no bar in the simulation, it is possible that the inner bins experience somewhat different dynamical conditions associated with the transition of the rotation curve from a more spherical to a disc-dominated potential. However, the affected radial bins are also those with the smallest number of stellar pairs, making it difficult for us to draw any strong conclusions about the origin of the enhanced decoherence in the inner disc. Overall, the lack of a strong radial trend is somewhat surprising, given the expectation of stronger perturbations in the inner disc.} However, it is important to note that we are measuring changes in action \textit{differences}, not the actions themselves. Hence, the result implies that conatal stars (that are born within 0.5 -- 2 pc of each other) experience similar perturbations regardless of their locations in the Galaxy, and thus their actions remain clustered to a comparable extent in both the inner and outer disc.

\section{Application to observations}
\label{sec:obs}

Having established a framework to quantify the divergence of stellar actions of pairs of stars as a function of initial separation and time, we now aim to apply these results to observed moving groups and stellar streams in the Milky Way disc. The goal of this section is to test whether present-day kinematic substructures are consistent with the hypothesis that they are unbound descendants of initially compact star clusters \citep{lada_lada2003,baumgardt_kroupa_2007,kuhn2019}, and to place quantitative constraints on their initial sizes using our simulation library.

The key diagnostic used throughout this section is present-day distribution of pairwise action differences within each candidate stream. Under the simplifying hypothesis that a stream formed in a single, compact star-formation event, all members share nearly identical actions at birth (i.e. zero pairwise action difference). As the system evolves in the Galactic potential for time $\tau$ (the present-day age of the object), action differences should grow according to the behaviour characterised by our simulation. Thus, we can infer the range of plausible birth separations for members of the stream.

Firstly, in \autoref{subsec:obs_data}, we summarise the data sources used and their preparation. We describe the methodology we use to compare the observed distributions with the simulation predictions in \autoref{subsec:obs_method}. In \autoref{subsec:obs_res_disc}, we present the results for the sample and discuss their implications. \autoref{subsec:completeness} talks about the completeness of observational membership and its effect on our analysis, while other caveats of the method are discussed in \autoref{subsec:caveats}.

\subsection{Data selection and preparation}
\label{subsec:obs_data}

Our observational sample is drawn from the catalogue complied by \citet{hunt2024}, which classifies objects previously identified as open clusters into bound clusters and unbound moving groups based on their Jacobi radii. 
The associated catalogue \citep{hunt_catalog} provides membership lists for all systems, along with per-member astrometry (RA, Dec, PMRA, PMDec, radial velocity and its uncertainty) from \textit{Gaia} DR3 \citep{gaiadr3} as well as cluster-level parameters such as age percentiles (16th, 50th and 84th) and current radius of the object in parsecs. We note here that \citet{hunt2024} do not consider radial velocities in their membership determination method. Hence, some stars with discrepant radial velocities may be miscategorised as members. To assess the robustness of our results to this effect, we repeated our entire analysis after excluding members that deviate by more than 3$\sigma$ in radial velocity from the group mean. This test, presented in the \aref{app:outliers}, yields similar results.

\begin{figure}
    \includegraphics[width=\columnwidth]{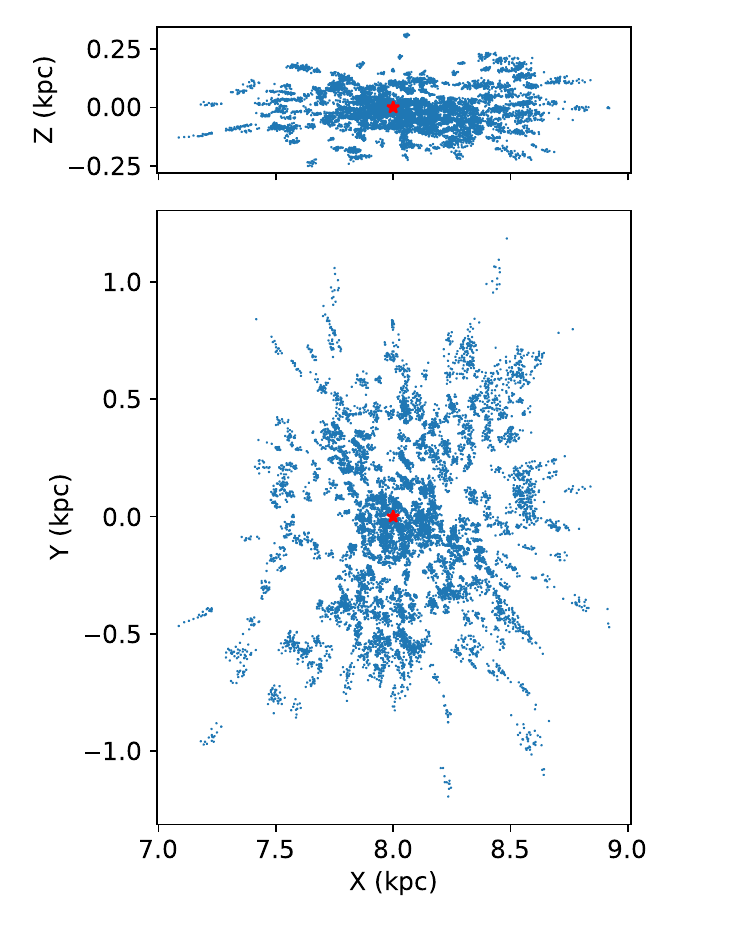}
    \caption{\edit{Spatial distribution of the stars used in our analysis in Galactocentric Cartesian coordinates. The position of the Sun is marked with a red star.}}
    \label{fig:objects}
\end{figure}

For all the members of each candidate moving group, we need to assemble their six-dimensional phase-space information for the computation of stellar actions. We adopt RA, Dec, proper motions and radial velocities directly from the membership list in the \citet{hunt_catalog}. When the catalogue does not provide a radial velocity, we supplement it using large spectroscopic surveys: GALAH DR4 \citep{galah}, APOGEE DR17 \citep{apogeedr17} and RAVE DR6 \citep{rave}, selecting the measurement with the smallest quoted uncertainty when multiple values are available. Distances are taken from the probabilistic parallax-based estimates of \citet{bailer-jones_dist}\edit{, which for our sample have a median uncertainty of 0.66\%}. For a small fraction of stars lacking an individual distance estimate, we assign the mean distance of the remaining members of the group as an approximation. This substitution is rare, since radial velocities are more often missing than distances, and stars without reliable velocities are removed from the sample altogether because accurate 6D phase-space information is essential for action computation. \edit{The spatial distribution of these objects in the Galactocentric Cartesian coordinates is shown in \autoref{fig:objects}, with the position of the Sun marked as a red star at the centre of the plot.}

\begin{figure}
    \includegraphics[width=\columnwidth]{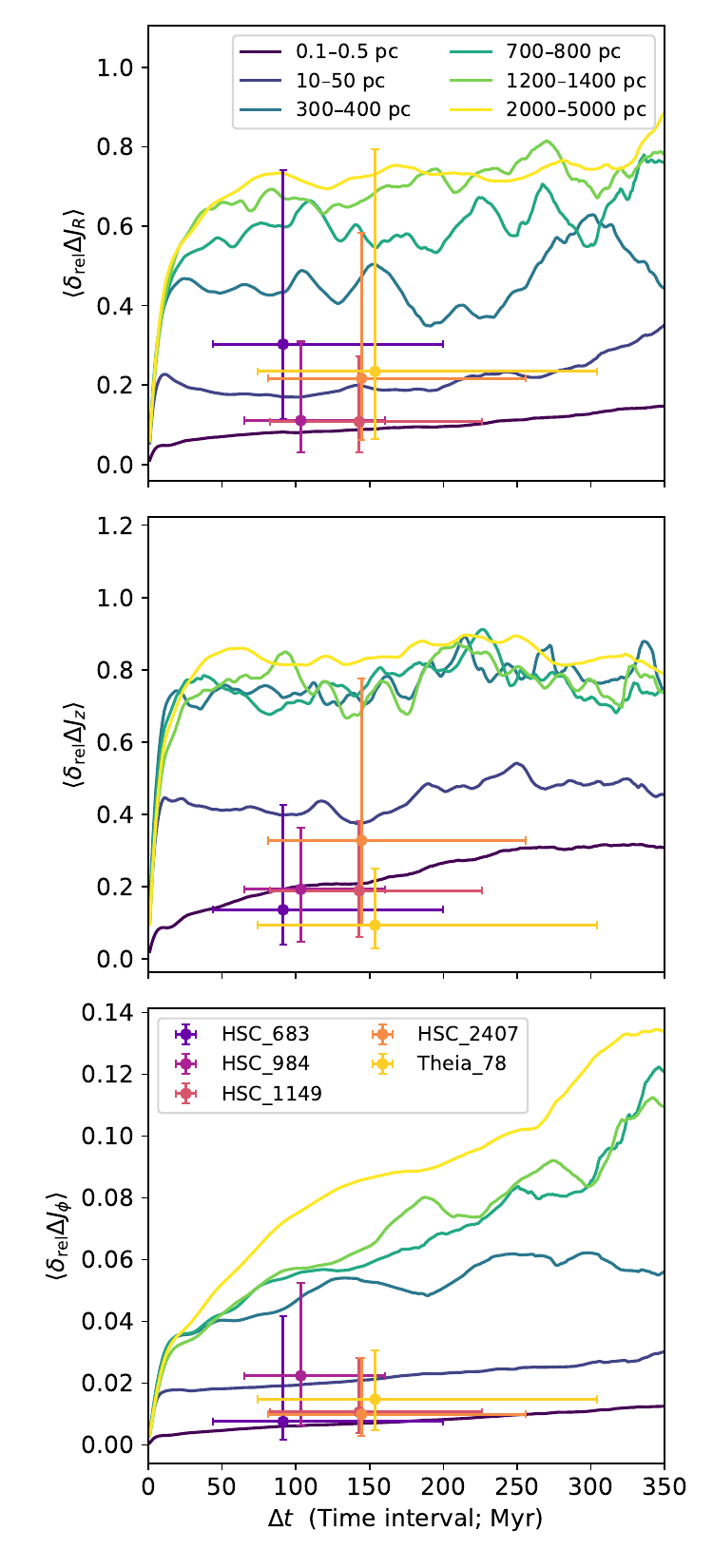}
    \caption{Relative action differences calculated for some of the observed moving groups (markers) placed on the theoretical grid of action-space evolution (coloured lines indicating different birth distance bins as shown in the legend). Rows from top to bottom show $\delta_\mathrm{rel}\Delta J_{R,z,\phi}$, respectively. Each point represents the median of the distribution $\{\delta\Delta J_\text{obs}\}$ and age of a moving group, with error bars showing the 16th and 84th percentiles \newedit{(see footnote 4)}. The lines are identical to those shown in \autoref{fig:rel_change}, though we show only a subset here to minimise clutter.}
    \label{fig:clusters}
\end{figure}

To ensure robust statistics for pairwise comparisons, we retain only those moving groups with at least ten member stars possessing full 6D information. This guards against unstable inference for poorly sampled systems. We also limit the sample to the parameter space covered by our simulations: objects are required to have an 84th percentile age less than 456 Myr, corresponding to the maximum $\Delta t$ in our simulated library. When the published age uncertainty is large, we retain the object provided that a significant portion of its age posterior lies within this range.

The actions of the member stars are calculated using the St\"ackel approximation \citep{binney2012} implemented in the \texttt{galpy} package \citep{bovy2015}. Since we have the 6D astrometry of the stars, we can directly integrate each star's orbit in a Milky Way potential which is taken to be the commonly-used \texttt{MWPotential2014} \citep{bovy2015} in our case. We integrate the orbits in this potential for 50 internal time units (corresponding to $\approx 1.9$ Gyr) and use the resulting orbit to estimate the local focal parameter $\Delta$ via the \texttt{estimateDeltaStaeckel} function. Given the potential and $\Delta$, we compute the actions using \texttt{actionAngleStaeckel}\footnote{We note that our simulation action predictions were computed using the epicyclic approximation described in \autoref{subsec: action calc}. However, tests in \citetalias{arunima2025} indicate that the epicyclic approximation and direct action computation produce comparable trends for the disc populations we consider (within $\sim 22\%$ for radial and $\sim 10\%$ for vertical action).}. 

For each moving group, we then take all possible pairs of stars and evaluate their relative change in action difference using \autoref{eq:rel_action} under the assumption that their initial action difference is zero, i.e. $|J_i(t) - J_j(t)| = 0$. Hence, for each stream, we get a distribution of observed relative action changes, represented by $\{\delta\Delta J_\text{obs}\}$, which characterises the present-day kinematic coherence of the stream. These values are upper limits, since if the initial action distances were non-zero then the value $\{\delta\Delta J_\text{obs}\}$ would be smaller.

Using the distribution $\{\delta\Delta J_\text{obs}\}$ and the system's age (and uncertainty), we can directly place each observed stream on the theoretical map of action-space coherence evolution shown in \autoref{fig:rel_change}. \autoref{fig:clusters} illustrates this comparison: coloured lines \newedit{indicate the median of change in action difference with respect to time for coeval pairs of stars, binned by their separation at birth, derived from the simulation (i.e. these curves are identical to those shown in \autoref{fig:rel_change})}. We then overplot a randomly-chosen subset of the observed moving groups as points, with the medians of their distribution $\{\delta\Delta J_\text{obs}\}$ on the vertical axis and median age on the horizontal axis; the error bars represent the 16th -- 84th percentile ranges in both quantities\footnote{\newedit{Vertical error bars reflect the spread or dispersion of the action differences, not the uncertainty on the median. In our analysis in \autoref{subsec:obs_method}, the full distribution of action differences is used in a Bayesian framework, so these error bars are purely illustrative. Horizontal error bars, on the other hand, represent the 16th -- 84th percentile age range as reported in \citet{hunt_catalog}, which is derived as a cluster-level parameter rather than individual stellar ages. When performing the inference of stream properties, we model the cluster age as a Gaussian distribution with mean equal to the catalogue median and standard deviation calculated using the 16th -- 84th percentiles. The posterior is then marginalised over this age distribution to correctly account for age uncertainty, ensuring the figure’s error bars do not bias the quantitative results.}}. The relative position of a moving group point with respect to the lines of different stellar birth separation provides an estimate of the upper limit on its initial size - the effective `birth radius' of the cluster that may have evolved into the present-day observed stream.

This qualitative comparison can be done for all the three components of action. However, visual inspection of \autoref{fig:clusters} shows that the vertical action yields the most stringent constraints on initial size. This is expected since, as seen in \autoref{subsec:decoherence} \edit{(and supported by \citealt{debattista2025})}, vertical action differences decorrelate over smaller physical scales than the azimuthal or radial components. Therefore, in the following sections, where we develop a quantitative, probabilistic inference framework for estimating initial cluster sizes, we restrict our analysis to the vertical action component.

\subsection{Methodology}
\label{subsec:obs_method}
Our goal is to estimate the most likely initial separation ($d_\text{init}$) of stars that make up present-day \edit{disc} streams by comparing their observed distribution of action differences to the simulation-based probability distributions of relative action difference changes, while marginalising over the uncertainty in the streams' age.

To explain our approach, we begin with the simplified case where we know the stream's age $\tau$ exactly, before generalising to the situation where $\tau$ is uncertain. Our first step for the simplified case is to construct the distribution of predicted action differences as a function of $d_\mathrm{init}$ for stars of this age based on our simulations. We therefore select all the stellar pairs whose age $\Delta t$ is equal to the stream age $\tau$, giving us a sample of pairs that are each characterized by an initial separation $d_\mathrm{init}$ and an increase in relative vertical action difference $\delta_\mathrm{rel} \Delta J_{z,\mathrm{sim}}$. We next select a random sample of 2 million pairs from the remaining data; we use 2 million rather than all the pairs ($\sim 10^8$ samples total for smaller values of $\Delta t$, diminishing for larger $\Delta t$) to keep the computational cost tractable. From these 2 million pairs, we construct a kernel density estimate (KDE) in the two variables $a \equiv \log d_\mathrm{init}$ and $x_\mathrm{sim} = \log \delta_\mathrm{rel}\Delta J_{z,\mathrm{sim}}$, using \edit{Silverman's rule to set the bandwidth of the Gaussian kernel}. We denote the resulting KDE in the two variables $\mathcal{P}_\text{2D}(x_{\text{sim}},a\mid \tau)$. We next reduce this to a series of 1D marginal KDEs for $x_\mathrm{sim}$ at each $a$ by evaluating
\begin{equation}
\label{eq:p1d}
    \mathcal{P}_\text{1D}(x \mid a,\tau) = \frac{\mathcal{P}_\text{2D}(x_{\text{sim}},a \mid \tau)}{\int_{-\infty}^{\infty} \mathcal{P}_\text{2D}(x_{\text{sim}},a \mid \tau)\, dx_{\text{sim}}}.
\end{equation}
In practice we compute this integral by evaluating the KDE on a 2D grid covering the full range of $a$ and $x_\mathrm{sim}$. After that, we extract slices at constant $a$, normalising the distribution at each $a$ to ensure that the total probability integrated over $x_\mathrm{sim}$ sums to 1 for each $a$; this normalization step corresponds to the denominator of \autoref{eq:p1d}. Finally, we construct an interpolating function for each value of $a$. The result is a function $\mathcal{P}_\text{1D}(x \mid a,\tau)$ that lets us efficiently predict the probability density function (PDF) of action differences $x$ we expect for a population born with a particular initial size $a$ and a present-day age $\tau$.

\begin{figure}
    \centering
    \includegraphics[width=\linewidth]{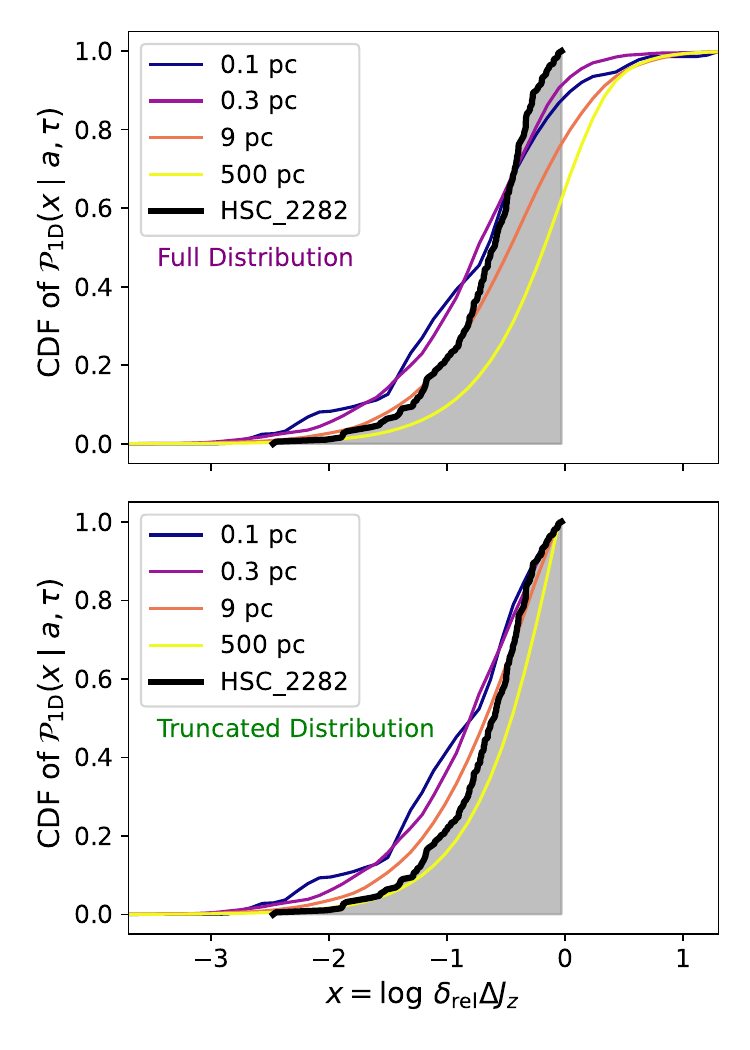}
    \caption{Coloured lines show examples of cumulative distributions corresponding to the one-dimensional probability distributions, $\mathcal{P}_\text{1D}(x\mid a,\tau)$, of log vertical action difference $x = \log\delta_\mathrm{rel}\Delta J_z$, for different initial separations $d_\text{init}$ (as indicated in the legend) and stellar populations of age $\tau=148$ Myr. The top and bottom panels show results for the full and truncated distributions, respectively -- see main text for details. The black solid line with shading shows the cumulative distribution for observed stellar pairs in the example stream HSC 2282 for comparison.}
    \label{fig:p1d}
\end{figure}

The top panel of \autoref{fig:p1d} shows examples of cumulative probability distribution functions (CDFs) corresponding to the PDFs $\mathcal{P}_\text{1D}(x \mid a,\tau)$ for different $a$ values and at an age $\tau = 148$ Myr, chosen to correspond to the central age estimate for an example stream, HSC 2282. The thick black line represents the CDF of log action differences $\{x_i\}$ for the observed members of the stream, while coloured lines represent predictions constructed from our simulations for different initial separations, as indicated in the legend. Notably, the observed CDF appears truncated near $x \sim 0$, and this behaviour is consistent across all streams we examined in our dataset, which suggests it could be an observational artefact: if stream members are too far away in action space (and, correspondingly, in physical space), they become increasingly unlikely to be observed or classified as stream members in the catalogue. To account for the possible effects of this observational limitation, we consider a modified method for constructing our theoretical PDFs $\mathcal{P}_\text{1D}(x \mid a,\tau)$ under the assumption that the lack of observed pairs at high $\delta_\mathrm{rel}\Delta J_z$ is primarily a result of observational incompleteness, in which case we must apply a similar truncation to the theoretical predictions. We therefore construct a modified version of $\mathcal{P}_\text{1D}(x \mid a,\tau)$, which we denote $\mathcal{P}_\text{1D, trunc}(x \mid a,\tau)$, by setting $\mathcal{P}_\text{1D}(x \mid a,\tau) = 0$ for $x>x_\text{max,obs}$ where $x_\text{max,obs}$ is the maximum observed value of $\log\delta_\mathrm{rel}\Delta J_z$ for a given stream, and renormalising the value of $\mathcal{P}_\text{1D}(x \mid a,\tau)$ at $x<x_\text{max,obs}$ so that the integral over all $x$ remains unity. We plot the CDF corresponding to this truncated distribution in the bottom panel of \autoref{fig:p1d}. In what follows we will carry out our remaining analysis using both the truncated and non-truncated versions of $\mathcal{P}_\text{1D}(x \mid a,\tau)$, since reality likely lies somewhere between the extreme assumptions that there are no observational biases against finding stream members with large action difference (corresponding to the top panel in \autoref{fig:p1d}) and that any disagreement between the observations and the predictions at high $x$ is solely due to observational bias (the lower panel).

Examining either panel of \autoref{fig:p1d}, it is clear that we can pick out by eye which values of $d_\mathrm{init}$ yield predict CDFs of action differences that are similar to the observed one. To render this intuition quantitative, we compute the likelihood of observing the action difference distribution $\{x_i\}$ given a particular pair of values $(a,\tau)$ as the product of individual probability for each observation:
\begin{equation}
\label{eq:likelihood}
    \mathcal{L}(\{x_i\}\mid a,\tau) =  \prod_{i} \mathcal{P}_\text{1D} (x_i \mid a,\tau)
\end{equation}
where $\mathcal{P}_\text{1D} (x\mid a,\tau)$ is the 1D PDF computed either for all the simulation data or for the truncated version. If we adopt a flat prior in $a = \log d_\text{init}$, the posterior probability distribution for $a$ then follows directly from the likelihood:
\begin{equation}
\label{eq:Ppost}
    \mathcal{P}_\text{post}(a \mid \{x_i\},\tau) \propto \mathcal{L}(\{x_i\}\mid a,\tau)
\end{equation}
This posterior describes the probability distribution for log initial size $a = \log d_\text{init}$ at a fixed age $\tau$. 

\begin{figure}
    \centering
    \includegraphics[width=\linewidth]{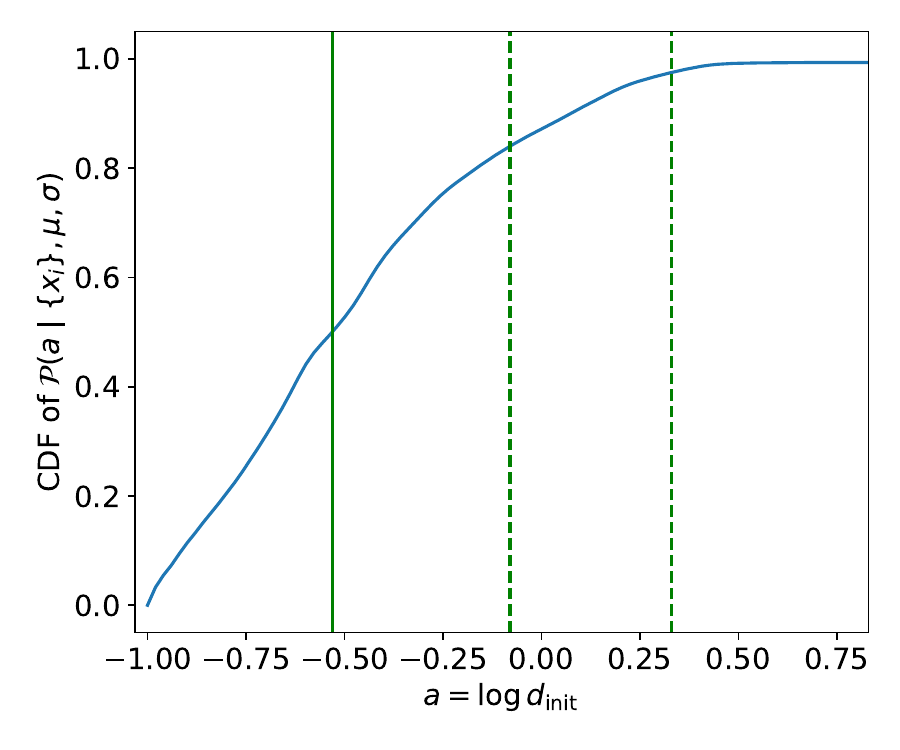}
    \caption{Cumulative distribution corresponding to the marginal posterior distribution, $\mathcal{P}(a\mid \{x_i\},\mu,\sigma)$, for HSC 2282, given its observed action differences $\{x_i\}$, age $\mu = 148.17$ Myr and $\sigma = 125.64$ Myr. This CDF is derived for the non-truncated assumption (see main text). The median logarithmic initial size $a_{50} = \log d_\text{init50}$ (in units of pc) is shown as the solid green vertical line while the 84th and 97th percentile upper limits $a_{84} = \log d_\text{init84}$ and $a_{97} = \log d_\text{init97}$ are shown as the two dashed green vertical lines to its right. }
    \label{fig:age marginalised- alessi}
\end{figure}

Now we can relax the assumption of known, fixed stream age $\tau$. Instead, we assume that the age is described by a Gaussian distribution with mean $\mu$ (the median age reported in the catalogue) and standard deviation $\sigma$ (derived from the 16th and 84th percentile ages assuming a normal distribution). To incorporate this observational uncertainty, we marginalise over $\tau$ by taking our final posterior PDF of $a$ to be
\begin{equation}
\label{eq:final_prob}
    \mathcal{P}(a \mid \{x_i\},\mu,\sigma) = \int \mathcal{L}(\{x_i\} \mid a,\tau) \mathcal{P}(\tau \mid \mu,\sigma)\, d\tau,
\end{equation}
where $\mathcal{P}(\tau \mid \mu,\sigma)$ is a Gaussian distribution with mean $\mu$ and dispersion $\sigma$. In practice, we perform this age marginalisation by drawing $\sim 10^4$ samples from the age distribution of the stream and evaluating conditional likelihood at each sampled age $\tau$ using \autoref{eq:likelihood}. Then, the likelihoods are averaged over the ensemble to obtain the age-marginalised posterior given by \autoref{eq:final_prob}. Again, we carry out this procedure for both the truncated and non-truncated versions of $\mathcal{P}_\mathrm{1D}$. For numerical stability, we restrict the sampled ages to $\tau \lesssim 397 $ Myr, since at later times the number of simulation pairs falls below $2\times10^6$, leading to noisy statistics as seen in \autoref{sec:results}. From the final posterior probability distribution $\mathcal{P}(a\mid\{x_i\},\mu,\sigma)$, we report the median ($d_\text{init50}$), 84th ($d_\text{init84}$), and 97.5th ($d_\text{init97}$) percentile, which provide a median estimate and upper confidence limits on $d_\text{init}$.\footnote{We do not report lower percentiles because, to remind readers, the values we obtain are always upper limits because they are derived under the assumption of negligible action separation at birth; any finite difference in the actions of stream members at birth would yield lower estimates of $d_\mathrm{init}$.} We show the CDF corresponding to the posterior for our example stream HSC 2282 in \autoref{fig:age marginalised- alessi}. 
    
Finally, to quantify how well the simulation and observed distributions agree, we compute the L$^1$ distance between them at each age sampled:
\begin{equation}
    \label{eq:l1}
        L^1(\tau) = \frac{1}{2} \int \left|\mathcal{P}_{\text{post}}(x \mid a_\text{ml},\tau) - \mathcal{P}_\text{obs}(x)\right|\, dx
\end{equation}
where $a_\text{ml}$ is the maximum-likelihood estimate of $a = \log d_\text{init}$. The factor of $1/2$ ensures that $L^1(\tau)$ ranges from 0 for identical distributions to 1 for completely non-overlapping ones. The final, age-marginalised value of $L^1$ for the stream is then:
\begin{equation}
        \label{eq:final_l1}
        L^1_f = \int L^1(\tau) \mathcal{P}(\tau\mid \mu,\sigma)\, d\tau
\end{equation}
This $L^1_f$ serves as a global diagnostic of how well the simulated distribution and its best fitting $d_\text{init}$ reproduces the observed one, with lower values correspond to better agreement.  As usual, we compute this diagnostic both using the full simulation data set and the version where we assume that the high $\delta_\mathrm{rel}\Delta J_z$ tail is truncated by observational selection effects.

\begin{figure}
    \centering
    \includegraphics[width=\linewidth]{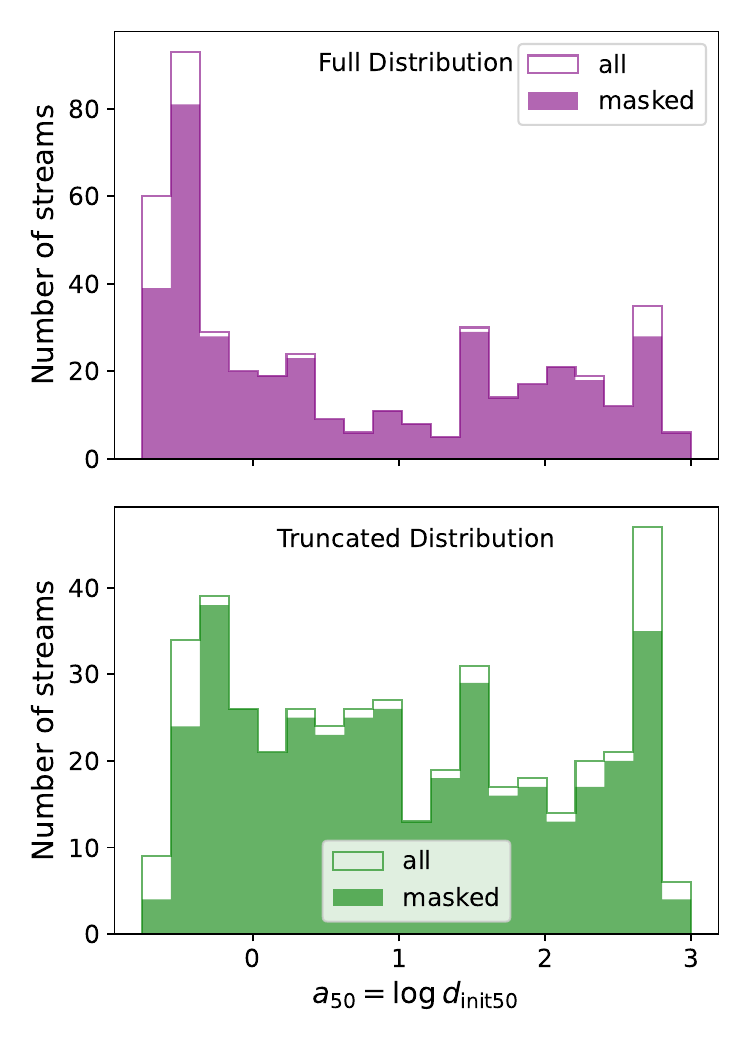}
    \caption{Distribution of the inferred median logarithmic initial sizes ($a_{50}$; in units of pc) for all streams in our sample. The shaded histogram highlights the subset of ``well-fit'' streams, selected by applying a cut on their $L^1_f$ values. The top panel shows results derived from the full posterior probability distribution, $\mathcal{P}(a\mid \{x_i\},\mu,\sigma)$, which is the appropriate distribution if the observed samples are complete, while the bottom panel shows those obtained from the truncated posterior distribution, which corresponds to the assumption that the observational samples are systematically missing stream members with large action differences.}
    \label{fig:d_init_dist}
\end{figure}

\begin{figure}
    \centering
    \includegraphics[width=\linewidth]{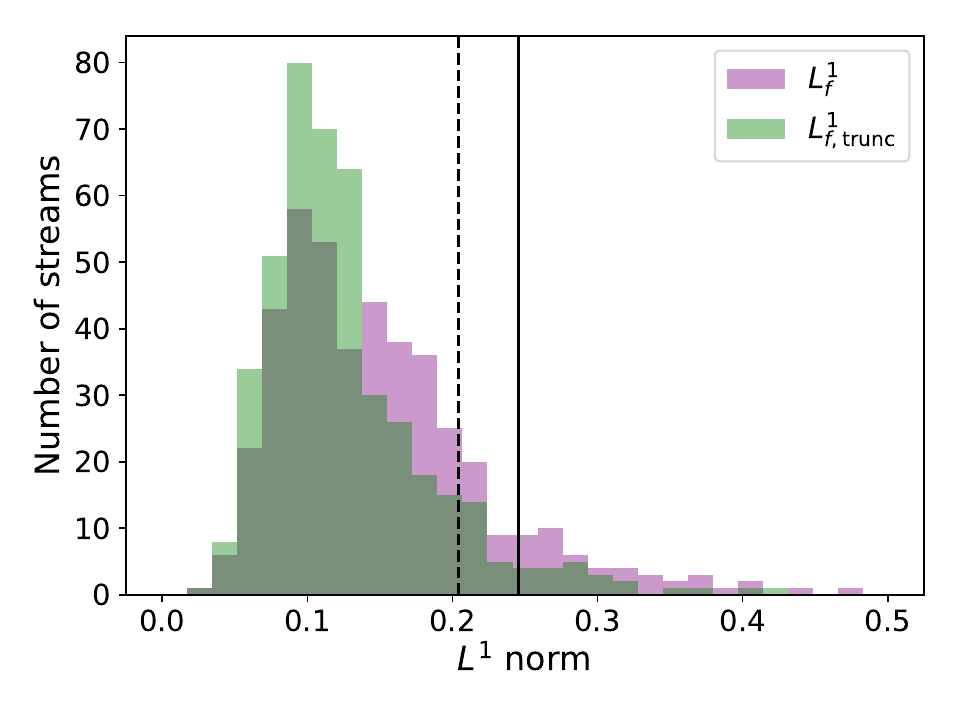}
    \caption{Distribution of the $L^1_f$ (purple) and $L^1_{f,\text{trunc}}$ (green) values that characterise how well the observed distribution of action differences matches the best-fitting simulation-predicted distribution. The black solid line represents the 90th percentile of $L^1_f$ while the dashed line shows the same for $L^1_{f,\text{trunc}}$.}
    \label{fig:l1_norm}
\end{figure}

\begin{figure}
    \centering
    \includegraphics[width=\linewidth]{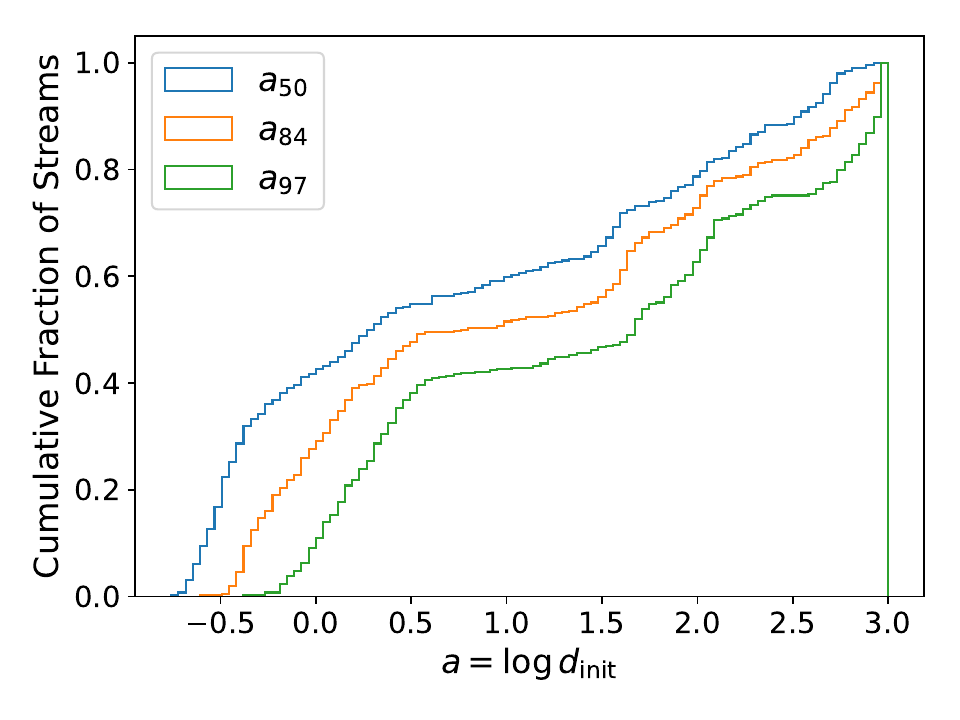}
    \caption{Cumulative distributions of the inferred logarithmic initial sizes (median $a_{50}$, 84th percentile $a_{84}$ and 97.5th percentile $a_{97}$) for the ``well-fit'' streams in our sample represented by the colours specified in the legend.}
    \label{fig:1_2sigma}
\end{figure}

\subsection{Results and discussion}
\label{subsec:obs_res_disc}

\begin{table*}
    \centering
    \begin{tabular}{c|c|c|c|c|c|c|c|c|c|c|c}
    \hline\hline
       Cluster Name  & Median age & Present size & $d_{\text{init}50}$ & $d_{\text{init}84}$ &$d_{\text{init}97}$ & $L^1_f$ & $d_{\text{trunc}50}$  & $d_{\text{trunc}84}$ &$d_{\text{trunc}97}$ & $L^1_{f\text{,trunc}}$ & $f_\text{comp}$ \\
       & [Myr] & [pc] & [pc] & [pc] & [pc] & & [pc] & [pc] & [pc] \\
    \hline
    ASCC\_100 & 64.89 & 21.17 & 0.28 & 0.42 & 0.65 & 0.19 & 0.38 & 0.70 & 1.42 & 0.16 & 0.91 \\
    ASCC\_104 & 121.70 & 32.19 & 476.77 & 871.23 & 983.36 & 0.16 & 493.33 & 885.88 & 985.08 & 0.18 & 0.98 \\
    ASCC\_18 & 9.26 & 16.17 & 0.45 & 0.62 & 3.21 & 0.09 & 0.63 & 0.85 & 3.24 & 0.10 & 0.95 \\
    ASCC\_33 & 215.38 & 21.42 & 0.31 & 0.73 & 1.86 & 0.17 & 0.83 & 2.97 & 14.56 & 0.14 & 0.85 \\
    ASCC\_69 & 72.90 & 26.99 & 0.25 & 0.48 & 0.99 & 0.15 & 0.92 & 2.92 & 13.87 & 0.11 & 0.82 \\
    ASCC\_73 & 158.29 & 38.93 & 0.32 & 1.34 & 2.83 & 0.18 & 29.21 & 49.36 & 73.43 & 0.08 & 0.85 \\
    Alessi-Teutsch\_4 & 132.97 & 28.06 & 0.20 & 0.35 & 0.62 & 0.32 & 0.48 & 1.20 & 3.11 & 0.16 & 0.67 \\
    Alessi\_13 & 24.75 & 42.07 & 659.41 & 819.64 & 983.16 & 0.06 & 666.91 & 879.90 & 989.71 & 0.06 & 0.99 \\
    Alessi\_145 & 168.03 & 82.18 & 0.30 & 0.45 & 0.74 & 0.27 & 0.53 & 0.94 & 1.71 & 0.14 & 0.84 \\
    Alessi\_84 & 117.08 & 27.70 & 0.42 & 0.74 & 1.45 & 0.14 & 0.48 & 0.88 & 1.57 & 0.15 & 0.99 \\
    \hline\hline
    \end{tabular}
    \caption{Summary of inferred parameters for the first 10 streams (alphabetically) in our sample. Columns list the stream name, median stellar age (in Myr), present-day physical size (in pc), inferred initial size percentiles from both the full and truncated posterior distributions (all in pc), associated $L^1$ norms quantifying the fit quality, and the completeness fraction $f_\text{comp}$. The stream name, median stellar age and present-day physical size are taken from \citet{hunt_catalog}. The full table is provided in machine-readable format (check Data Availability).}
    \label{tab:table1}
\end{table*}

\autoref{fig:d_init_dist} presents the distribution of inferred median initial sizes ($a_{50}$) for all the streams in our sample. The top panel shows the values obtained from the full posterior probability distribution that corresponds to assuming no observational incompleteness, while the bottom panel shows the results derived from the truncated version of the distribution that corresponds to assuming that all high $\delta_\mathrm{rel}\Delta J_z$ stars have been missed. \autoref{fig:l1_norm} displays the distribution of the corresponding $L^1_f$ and $L^1_{f,\text{trunc}}$ values for all streams, which quantify the quality of the fit -- smaller values indicate better agreement between the observed and modelled action-difference distributions. Most streams exhibit $L^1 \lesssim 0.2$, though the distribution shows a mild tail toward poorer fits (still $<0.5$). We mark the 90th percentiles of the two distributions with vertical black lines (solid for $L^1_f$ and dashed for $L^1_{f,\text{trunc}}$), which effectively delineate the tail region. Streams to the left of this line are considered well-fit by our model. The filled histograms in the top and bottom panels of \autoref{fig:d_init_dist} show the median initial size distributions for these ``well-fit'' streams, while the unfilled histograms show the results if we do not omit these less well-fit streams; the overall shape of the distribution remains largely unchanged after this quality cut. To illustrate the associated uncertainties in the inferred sizes, \autoref{fig:1_2sigma} shows the CDFs of median initial size ($a_{50}$) along with the 84th and 97.5th percentile initial size ($a_{84}, a_{97}$) for the entire sample.

All derived quantities -- including the median, 84th, and 97.5th percentiles of inferred initial sizes, corresponding $L^1_f$ statistics (for both the full and truncated probability distributions), stream name, and its median stellar age -- are compiled in \autoref{tab:table1}. We also provide an estimated completeness fraction, $f_\text{comp}$, of each stream's membership under the hypothesis that high action difference stream members have been missed -- see \autoref{subsec:completeness} for full details of how we compute this quantity. The complete table is available in a machine-readable format as supplementary material.

Examining the distribution of the inferred initial sizes in \autoref{fig:d_init_dist}, we find that the streams span more than three orders of magnitude in $d_\text{init50}$. We note that these streams have an average present size of $\sim 30$ pc. The fact that most inferred birth sizes are tens to hundreds of times smaller suggests that a large fraction of these systems were compact stellar clusters that have since become unbound and dispersed along their orbits. We interpret the left side of the distribution, where $d_\text{init50} \lesssim 3$ pc, as representing these compactly born objects, whose members have gradually phase-mixed to produce the extended kinematic streams observed today. In contrast, the right side of the distribution contains objects with inferred initial sizes of several hundreds to thousands of parsecs, comparable to or larger than their current size. Such systems could not plausibly have been bound clusters in the past. Instead, we propose that these groups were formed through resonant processes, in which stars are trapped or aligned in action space by non-axisymmetric perturbation in the Galactic potential. These mechanisms have been associated with spiral arms and bar resonances \citep[e.g.,][]{sellwood2010,khoperskov2020,trick2021,palicio2023}, which can generate coherent moving groups without a common birth origin \citep{bar_movinggroups25}. The action difference distributions of these systems resemble those of randomly selected stars in our simulation (where correlations in action differences have already decohered; see \autoref{fig:abs_change}), consistent with a non-coeval, resonance-driven origin. Another possibility is that some of these groups with large initial birth size consist of stars that formed in the same large-scale complex \edit{or with high initial velocity dispersion}, leading to an intrinsic spread in birth actions that mimics a coherent kinematic substructure at the present day. 

\begin{figure}
    \centering
    \includegraphics[width=\linewidth]{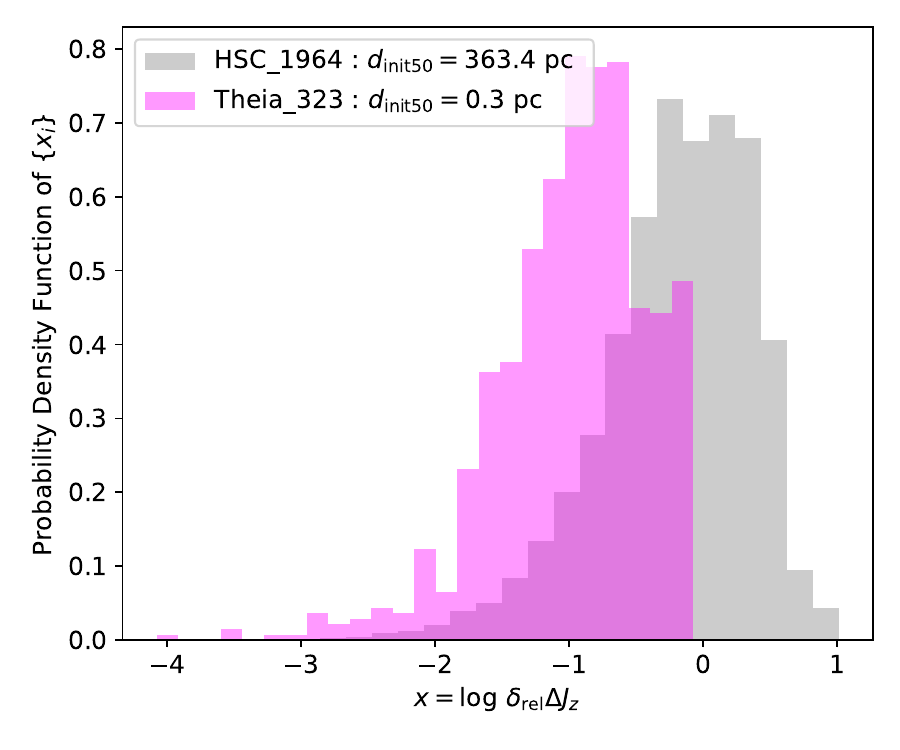}
    \caption{The observed pairwise action differences logarithm $\{x_i\}$ distribution for two representative streams: (i) Theia 323 (in pink) with inferred initial size $d_\text{init50} = 0.3$ pc, indicative of a compact origin stream; and (ii) HSC 1964 (in grey) with inferred initial size $d_\text{init50} = 363.4$ pc, representative of a resonant structure.}
    \label{fig:jobs_comp}
\end{figure}

In order to demonstrate that our analysis method is picking up on real differences in stellar properties when assigning some streams small $d_\mathrm{init}$ and others much larger values, in \autoref{fig:jobs_comp} we provide a representative comparison between the observed action-difference distributions for two stream cases: Theia 323, with inferred birth size of 0.3 pc, and HSC 1964, which has an inferred birth size of 363.4 pc. It is clear from the figure that Theia 323 shows visibly lower pair-wise action differences than HSC 1964. This validates our Bayesian inference framework, demonstrating that it is capturing real variations in the action-space coherence of stellar groups. 

\begin{figure}
     \centering
     \includegraphics[width=\linewidth]{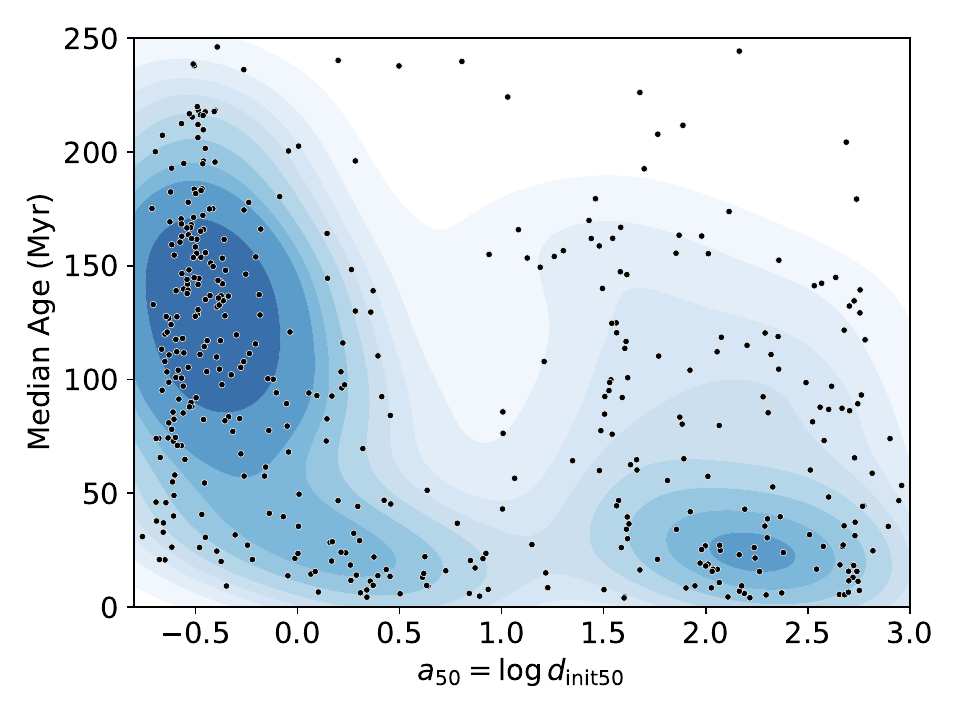}
     \caption{Distribution of the inferred median logarithmic initial sizes ($a_{50}$) for all streams in our sample, plotted as a function of their median stellar ages (black points). Blue contours indicate regions of higher density in the $(a_{50}, \text{age})$ plane, highlighting the overall structure of the distribution.}
     \label{fig:age_dep}
 \end{figure}

We further examine whether the inferred initial sizes depend systematically on the age of the stream. In \autoref{fig:age_dep}, we plot the joint distribution of $a_{50} = \log d_{\text{init}}$ and median stellar age for each stream, using age estimates from the \cite{hunt_catalog} catalogue. We find that streams identified as disrupting clusters (those lying to the left of the figure) span the full age range of our sample, with no obvious bias to young or old age. Those with large inferred initial separations based on their action difference distribution are mildly biased to younger ages, but not by a very significant amount. \edit{If these objects share a common origin as initially extended or dynamically hot structures, this trend can be understood as a selection effect: as such dynamically-hot objects age, they will be less likely to remain as coherent structures that could be identified in the present-day data, and thus would be selected against. Alternatively, if these objects originate from Galactic resonances, the mild bias to the younger ages can be explained as follows: dynamically cold populations are more susceptible to being captured by resonances, while older populations having already been heated by secular processes and perturbations are less likely to retain a coherent signature of being trapped in a resonance.}

\subsection{Completeness of the data}
\label{subsec:completeness}

If we adopt the hypothesis that the distribution of action differences $\delta_\mathrm{rel}\Delta J_z$ is truncated at the high end because the stream membership lists are incomplete, we can use the comparison between the observed distributions and our best-fitting models to estimate what fraction of stream members have been missed. We quantify this by for each stream defining the completeness fraction, $f_\text{comp}$, which we compute by comparing the CDF of observed action differences for the stream members with that of a non-truncated probability distribution $\mathcal{P}_\text{1D}(x \mid a,\tau)$, which represents the expected distribution in the absence of incompleteness/truncation. Here we take $x=\text{max}(x_i)$ -- the logarithm of largest observed action difference, and $a = d_\text{trunc50}$ -- the inferred initial size of the stream assuming truncation. The key idea is that the non-truncated probability distribution derived from our simulation data, $\mathcal{P}_\text{1D}(x \mid a,\tau)$), is considered the ``true" action difference distribution. However, due to incompleteness in observational data, we can not directly observe this distribution. Instead, we use the truncated model to estimate the best-fit initial size, $d_\text{trunc50}$, and  use this to assess how the true action difference distribution should appear. By comparing this theoretical, non-truncated distribution to the observed distribution, we can determine the fraction of stream members that are likely missing.
For example, in the case of HSC 2282, the best-fit initial size is found to be $d_\mathrm{trunc50} \approx 9$ pc. As shown in the upper panel of \autoref{fig:p1d}, the observed CDF reaches $\approx 75\%$ of the model CDF for $d_\text{trunc50} = 9$ pc (orange curve). This implies that the stream membership is about 75\% complete, with the remaining 25\% of members likely missing from the current catalogue. We emphasise that this is likely an upper limit on the true level of incompleteness, and a lower limit on the completeness, since it is derived under the assumption that \textit{all} of the discrepancy between observed and modelled action differences at high $\delta_\mathrm{rel}\Delta J_z$ is due to incompleteness.

\begin{figure}
     \centering
     \includegraphics[width=\linewidth]{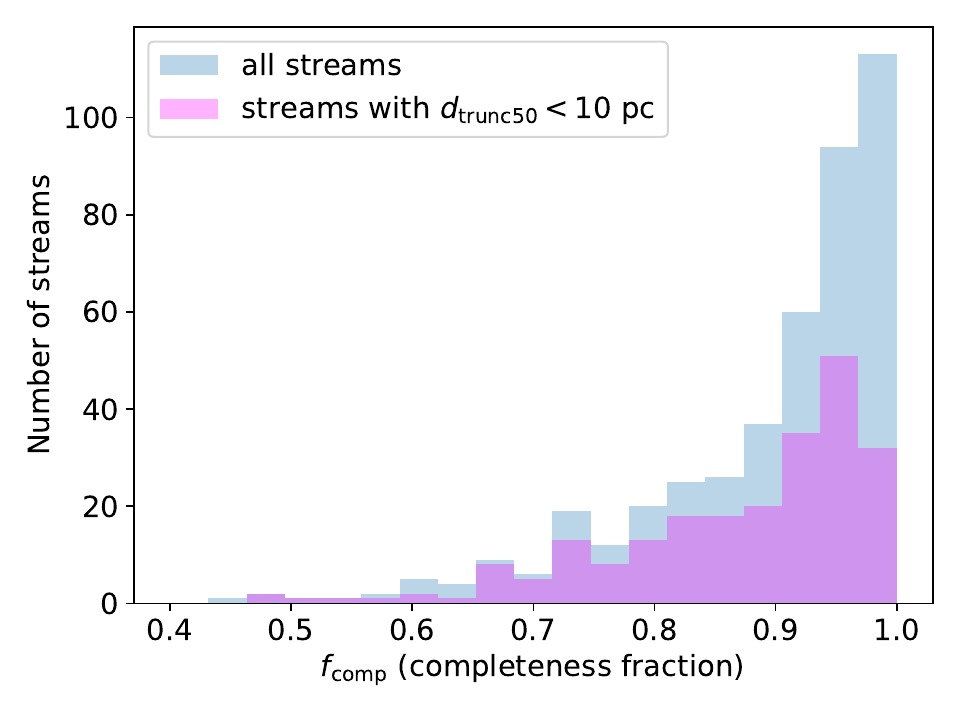}
     \caption{Distribution of the completeness fraction ($f_\text{comp}$) for all streams in our sample shown in blue. The pink histogram shows the same only for streams whose inferred initial size  $d_\text{trunc50} <10$ pc so they likely began as compact clusters instead of the probable resonant structures. This separation is useful to explore whether the streams that are more likely to have formed as compact clusters exhibit different completeness characteristics. }
     \label{fig:completeness}
 \end{figure}

We show the distribution of the completeness fractions for all streams calculated using this approach in \autoref{fig:completeness}. The blue histogram represents $f_\text{comp}$ for all the streams in our sample. However, it may not be meaningful to assess the `completeness' of \edit{streams with large inferred initial sizes,} since they most likely do not share a common origin. To address this, we also show the $f_\text{comp}$ distribution for streams with inferred sizes $d_\text{trunc50}<10$ pc, which are likely to have been compact clusters in the past. The distribution for this subset of streams is shown in pink in the figure. We find that only a small fraction of members are missing. The median $f_\text{comp}$ for all streams is 0.93, and for streams with $d_\text{trunc50}<10$ pc, it is 0.91. Thus, about 10\% of the stream members are likely missing on average, with these missing members most likely concentrated in the high-tail end of the action difference distribution.

While this analysis suggests that incompleteness is modest, its effects are significant. Comparison between the top and bottom panels of \autoref{fig:d_init_dist} highlights how missing data can significantly affect the inferred size distribution of the streams. In the non-truncated case (top panel), we assume that none of the members are missing and that there is no high-tail end in the observed action differences. This results in most streams having small inferred initial sizes. However, when we account for the possibility that the observational membership is incomplete and truncate our simulation model distribution at the highest observed action difference value, we obtain systematically larger initial sizes, as shown in the bottom panel. The non-negligible difference in the distributions demonstrates that the presence or absence of the high-action-difference tail can have a significant effect on the inferred initial size. If these members exist but are not detected, the true birth size of the stream must be larger to account for the missing high-action-difference tail.

This implies that if we wish to improve our ability to differentiate between \edit{disc streams that represent large birth scale populations} and those that represent dispersion clusters, it is critical to improve, or at least better characterise, the completeness of stream membership lists. The improvement is most needed for high-action-difference stars whose kinematic properties may place them at the edges of the observed action-space distribution, potentially leading to their exclusion from stream catalogues.
One possible approach to this problem would be to expand the search in chemical space, by using the chemical properties of stars such as elemental abundances to identify stream members. The use of chemical abundances in addition to kinematic data is becoming increasingly feasible with large, high-resolution spectroscopic surveys like the APOGEE survey \citep{apogeedr17}, and GALAH \citep{galah} which provide precise elemental abundances for stars across the Milky Way. By combining both action and chemical space, we may be able to identify members that would otherwise be missed in a search based solely on position and kinematic information.


\subsection{Caveats}
\label{subsec:caveats}
Several caveats must be considered when interpreting these results. First, since we specifically chose `unbound' stars in the simulation analysis, our results are based on the assumption that all stream members have been unbound for their entire lifetime. While the \citet{hunt_catalog} catalogue classifies these groups as `moving groups' (and unbound) based on their Jacobi radii, many would have been bound in the past for a fraction of their age. Strictly speaking, the relevant age for our model would then be the time since they became unbound. However, given the large uncertainties in age estimates (typically $\gtrsim 30\%$), this distinction has limited practical impact.

Second, the inference framework assumes that our simulation library is based in the relevant range of the Galactic environment in which the streams we are applying it to have been. The simulation samples the full Galactic disc, providing a population-averaged model. Consequently, applying these results to a specific Galactic region requires caution. Young clusters residing in highly perturbed environment (e.g., near the bar, or in a tidally disrupted area) may experience stronger diffusion than is implied by our global calibration. However, older systems, which have orbited the Galaxy multiple times and sampled diverse environments, are expected to be better represented by our population-averaged model.

Finally, the reliability of our inferred quantities ultimately depends on the accuracy of stream membership, radial velocity measurements (which are often absent or have large uncertainties), the adopted Galactic potential model, and the approximations involved in the computation of actions. The improved astrometry and increased availability of radial velocity in the upcoming \textit{Gaia} DR4 will mitigate some of these limitations. A more comprehensive, error-inclusive analysis that propagates the observational uncertainties is planned for future work. 

It is also important to re-emphasise that our inferred initial size estimates represent upper limits. The model assumes that the group members were initially co-located in action space (i.e. had negligible initial $\Delta J$). If the true birth action spread were non-zero, a more compact initial configuration would be required to reproduce the same observed action differences in the present. In addition, as mentioned in \autoref{subsec:simulation}, we may be overestimating the rate of action decoherence due to the resolution of our simulation's star particles. This, too, biases our inferred initial sizes towards an upper limit.
  

Overall, our findings support a dichotomy in the nearby kinematic substructures: (i) disrupted star clusters that have evolved into coherent stellar streams, and (ii) resonant or dynamically induced moving groups with no common formation origin. 
This method provides a new pathway to constrain the origins of streams, complementing chemical tagging and orbit integration approaches, and lays the groundwork for systematic application to the growing catalogue of streams uncovered by \textit{Gaia}.



\section{Conclusions}
\label{sec:conclusions}
In this paper, we use high-resolution MHD simulation of a Milky Way-like galaxy to examine how stars that form together evolve in action space within a fully self-consistent environment that includes gas dynamics, star formation, and a live stellar disc with flocculent spiral arms. The main outcomes of our study are as follows:

\begin{itemize}
    \item Stars born in close proximity to each other maintain strongly correlated actions throughout the $\sim 0.5$ Gyr period of our simulation. Their action differences show much smaller change than those of stars born farther apart, remaining well below 50\% in relative change. In contrast, pairs separated by larger distances at birth show uncorrelated evolution, consistent with the stochastic diffusion seen for individual stars.
    \item The spatial scale over which co-natal stars lose correlation in action space varies by action component. Vertical action coherence breaks down over several hundred parsecs, while radial and azimuthal components become decorrelated only on kiloparsec scales. These differences reflect the dominant physical drivers of action evolution: vertical actions are perturbed by local structures such as GMCs (vertical disc scale height is of the order of hundred parsecs), while radial and azimuthal actions are shaped by large-scale non-axisymmetric features such as spiral arms which operate on kiloparsec scales.
    \item We find no \edit{strong} dependence of the degree of action-space coherence on Galactocentric radius. \edit{Across most of the disc, s}tars born close together in both the inner and outer disc experience similar levels of correlated perturbation, implying that different processes perturb their actions collectively rather than differentially i.e. they are moving together in logarithmic action space rather than being spread apart. \edit{A modest enhancement in decoherence is observed in the innermost radial bins, particularly at late times; however, these regions are strongly affected by limited statistics preventing us from drawing firm conclusions about the underlying physical origin of this behaviour.}
   \item We construct a probabilistic inference framework that links the present-day action distributions of \edit{disc} streams to their birth scales, using the library of action-difference evolution from the simulation. This allows the typical initial size of a stream to be estimated from its observed phase-space structure and age. Applying this framework to the \citet{hunt2024} catalogue, we derive initial size estimates for 438 moving groups and quantify their fit quality and membership completeness. We provide these results as a resource for the community.
    \item The inferred birth sizes suggest a dichotomy among the \edit{disc} streams. Most systems likely were in the past compact clusters that have since dispersed into streams, while others appear too extended to have been bound and are better explained by resonant or dynamically induced structures -- possibly linked to the Galactic bar, spiral arms, or large star-forming complexes -- rather than by common birth origin.  
\end{itemize}

Together, these results establish the characteristic spatial and temporal scales over which co-natal stars lose dynamical coherence in the realistic Galactic environment of our high-resolution MHD simulation. By explicitly connecting present-day action-space structure to stellar birth separations, we provide a quantitative calibration of how clustered star formation 
maps onto the evolving phase-space distribution of disc stars. 
This framework provides a method for co-natal tagging, complementing chemical-abundance and orbital methods in the Milky Way disc.
The advent of improved astrometry and radial velocity coverage with 
\textit{Gaia} DR4 and future spectroscopic surveys  will make such comparisons increasingly powerful.  Incorporating observational uncertainties into our inference framework,  and extending the simulation suite to include additional large-scale perturbations such as the Galactic bar will make this approach directly applicable to real data. 

\section*{Acknowledgements}

This research was undertaken with the assistance of resources from the National Computational Infrastructure (NCI Australia) and the Pawsey Supercomputing Centre, NCRIS-enabled capabilities supported by the Australian Government, through award jh2. AA and MRK acknowledge support from the Australian Research Council through Laureate Fellowship FL220100020.

\section*{Data Availability}

The codes and some of the data used in this study are available at \url{https://github.com/aruuniima/stellar-actions-II}, along with the complete version of \autoref{tab:table1}. The entire simulation outputs are not included in the public repository due to their size, but will be provided by the authors upon reasonable request.



\bibliographystyle{mnras}
\bibliography{example} 




\appendix

\section{\edit{Dependence on neighbour number}}
\label{app:NN}

\begin{figure}
    \centering
    \includegraphics[width=\linewidth]{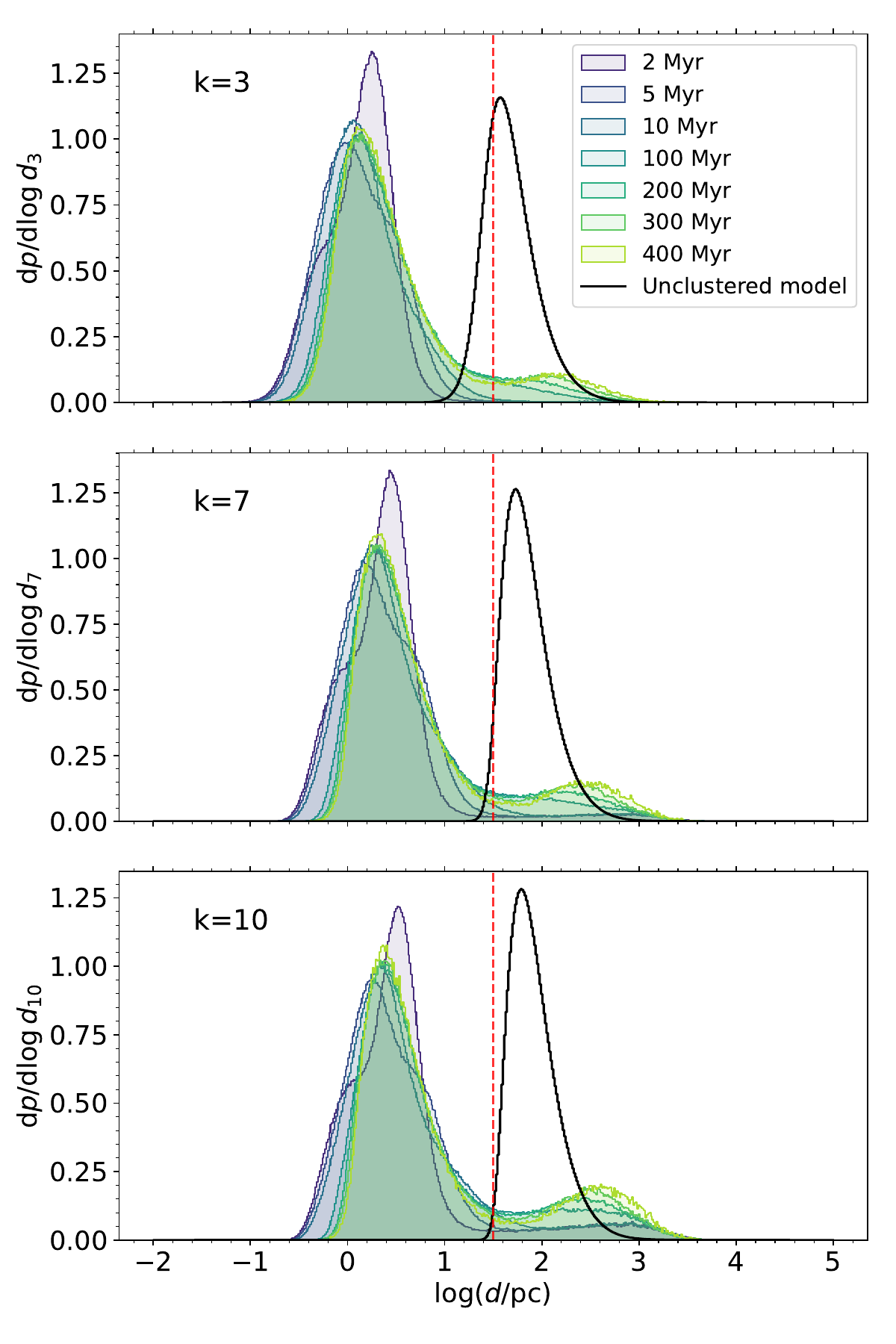}
    \caption{\edit{Same as \autoref{fig:NNdist}, but now showing distribution of (in order from top to bottom panel) 3rd, 7th and 10th nearest neighbour distances for coeval stars at different ages. The distribution of the same distance for the star population sampled from a double exponential disc model fit to our simulation is shown as the black curves (scaled for ease of visual comparison). We show the 30 pc limit that we use to separate `clustered' stars from dispersed stars as the red dashed vertical line.}}
    \label{fig:NN_alt}
\end{figure}

In \autoref{subsec:NNsec}, we use the distribution of the 5th nearest neighbour to identify stars that have dispersed from their natal clusters. In this appendix, we verify that our results are not sensitive to the specific choice of 5th nearest neighbour rather than some other number. In \autoref{fig:NN_alt} we plot the distribution of neighbour distances computed using the 3rd, 7th, and 10th nearest neighbours, both for our simulation results and for our control unclustered sample; this figure may be compared directly to \autoref{fig:NNdist} of the main text, which is computed for the 5th nearest. The figure shows that adopting neighbour numbers from $k = 3$ to 10 yields results that are qualitatively very similar to using the 5th nearest, including the emergence of a bimodal separation distribution at later ages. In all cases, the location of the minimum separating the clustered and dispersed populations depends only weakly on the number of neighbours, $k$.

\newedit{We do observe that the control distribution (the black curve) shows greater sensitivity to the neighbour number $k$. This arises from the different ways nearest-neighbour statistics sample point sets with and without clustering. In an unclustered population such as the control distribution drawn from a smooth background density, higher values of $k$ correspond to progressively larger spheres enclosing additional points, so the $k$-th nearest neighbour distance distribution shifts more markedly with $k$. For a clustered population, on the other hand, the closest neighbour up to a moderate value of $k$ will all sample the same compact structure, so the first peak of this distribution, which identifies the cluster scale, changes very little with $k$. The second peak of the clustered simulation distribution, corresponding to stars that have dispersed beyond the cluster scale, is naturally broader and occurs at larger distances. It varies with $k$ because increasing $k$ will include more distant stars when computing the $k$-th nearest neighbour. }

\newedit{Increasing $k$ tends to reduce variance in the neighbour-distance estimate and narrow the distributions, making features such as the separation of peaks more pronounced, but may also introduce smoothing bias if $k$ is too large. This bias-variance trade-off is well recognised in the literature \citep{beraldoesilva2025} and motivates our choice of $k=5$ as a conservative compromise that balances sensitivity to genuine clustering with stability against noise. }

\section{\edit{Testing alternative threshold separations}}
\label{app:threshold}
\begin{figure}
    \centering
    \includegraphics[width=\linewidth]{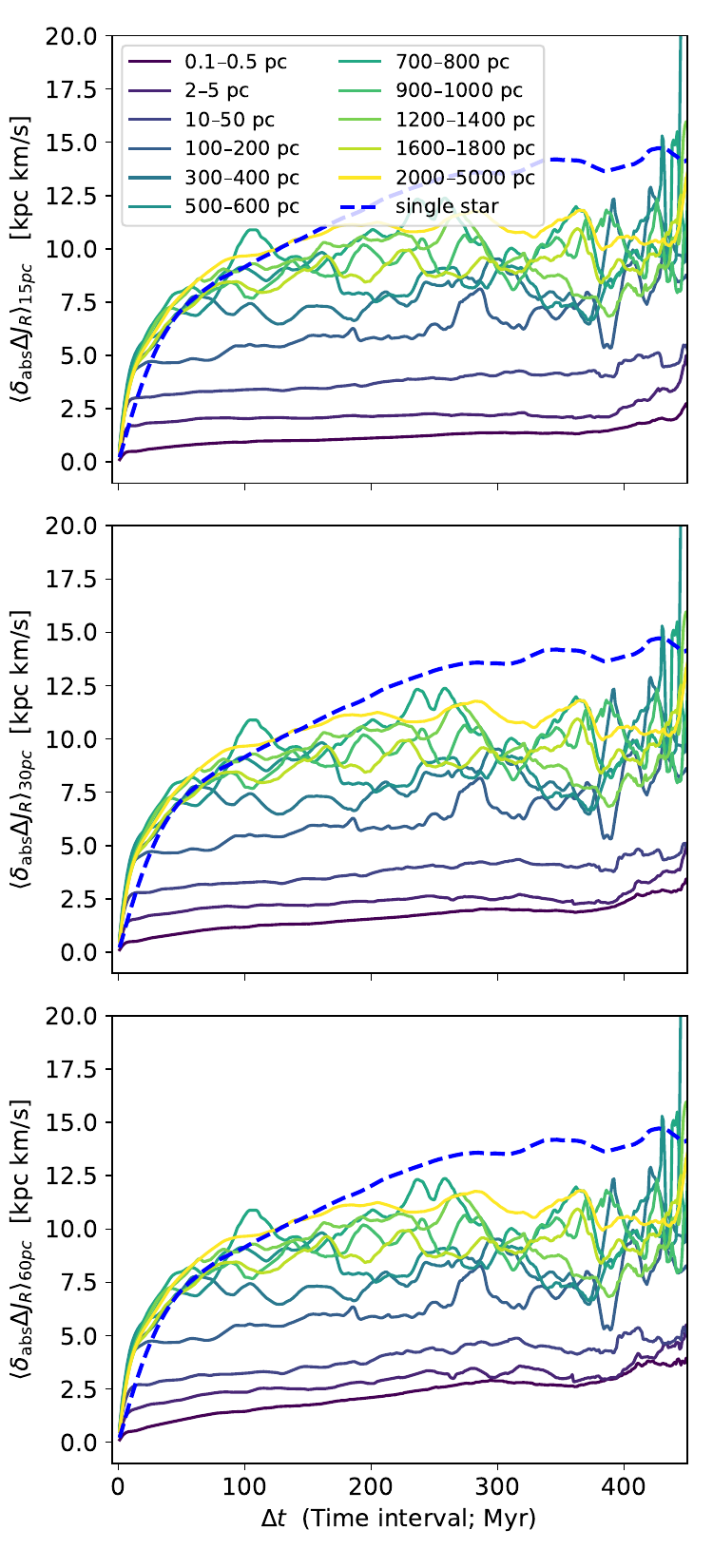}
    \caption{\edit{Same as the first panel of \autoref{fig:abs_change} but with different sample selection criteria: the top panel shows results where we retain in our sample stellar pairs separated by at least 15 pc at 100 Myr, the middle panel shows our fiducial value of 30 pc (and is therefore identical to the top panel of \autoref{fig:abs_change}), and the bottom panel shows results for 60 pc.}}
    \label{fig:JR_threshold}
\end{figure}

\begin{figure}
    \centering
    \includegraphics[width=\linewidth]{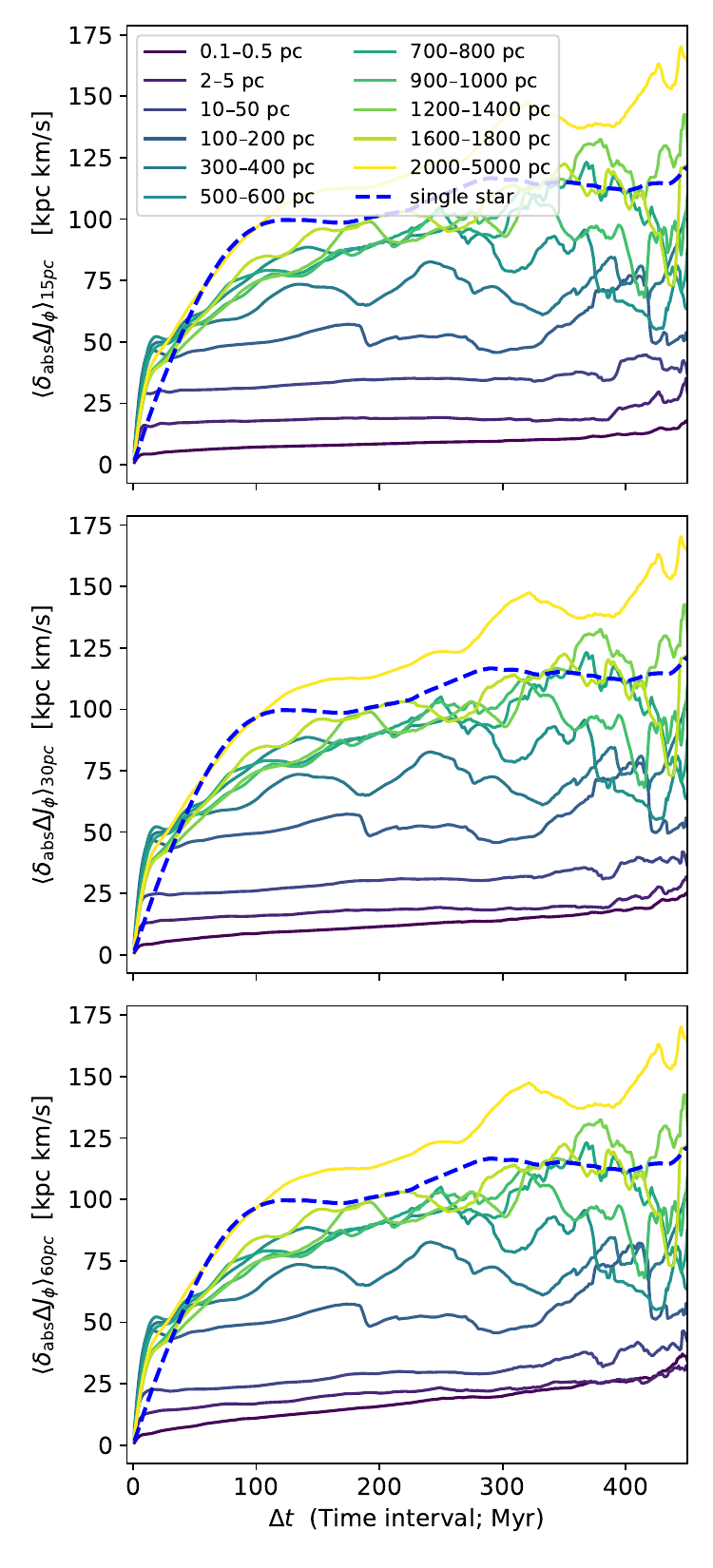}
    \caption{\edit{Same as \autoref{fig:JR_threshold} but for azimuthal action difference change, $\delta_\text{abs}\Delta J_{\phi}$.}}
    \label{fig:Jphi_threshold}
\end{figure}

\begin{figure}
    \centering
    \includegraphics[width=\linewidth]{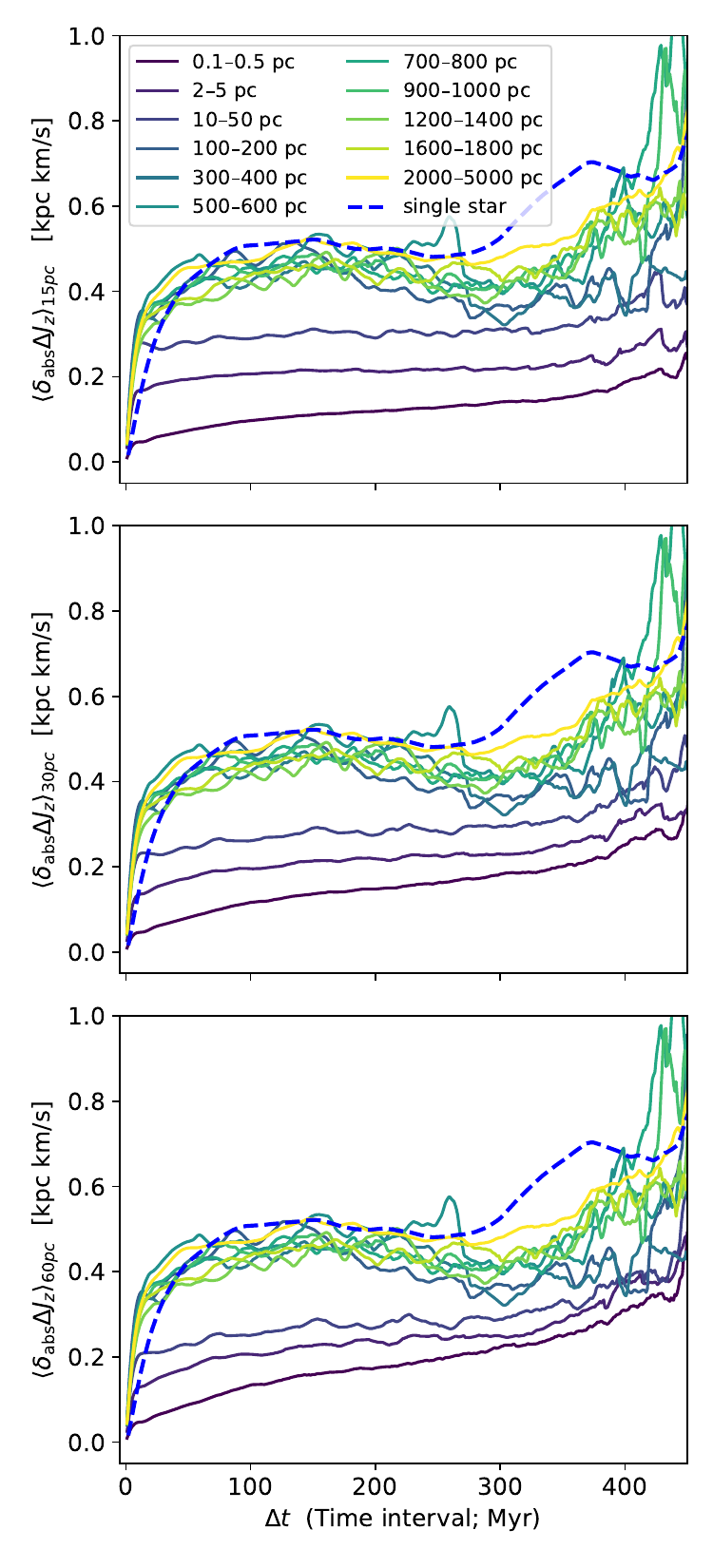}
    \caption{\edit{Same as \autoref{fig:JR_threshold} but for vertical action difference change, $\delta_\text{abs}\Delta J_{z}$.}}
    \label{fig:Jz_threshold}
\end{figure}

\edit{In the main text, we define our sample of stars that have drifted apart, on which we base our analysis, by requiring that stellar pairs must have a separation greater than $30$ pc at $\Delta t = 100$ Myr after birth. Here, we test the sensitivity of our results to this choice by repeating the analysis in our paper using alternative threshold separations of 15 pc and 60 pc (i.e., a factor of two smaller and larger than our fiducial value). We show the median absolute change in action difference for radial, azimuthal and vertical components \autoref{fig:JR_threshold}--\autoref{fig:Jz_threshold}, with increasing threshold separation -- 15 pc in the top panels, 30 pc in the middle panels, and 60 pc in the bottom panels. The middle panels here are identical to the results shown in \autoref{fig:abs_change}, but we repeat them here for easy comparison. The figures demonstrate that changing our criteria for separated stars by factors of two in either direction does not qualitatively affect any of the results presented in this work -- indeed, the different panels are almost indistinguishable. We therefore retain the 30 pc threshold for the selection of the stellar pairs used in the main analysis. }

\section{\edit{Testing the effect of potential outliers in membership lists of objects}}
\label{app:outliers}

\begin{figure}
    \centering
    \includegraphics[width=\linewidth]{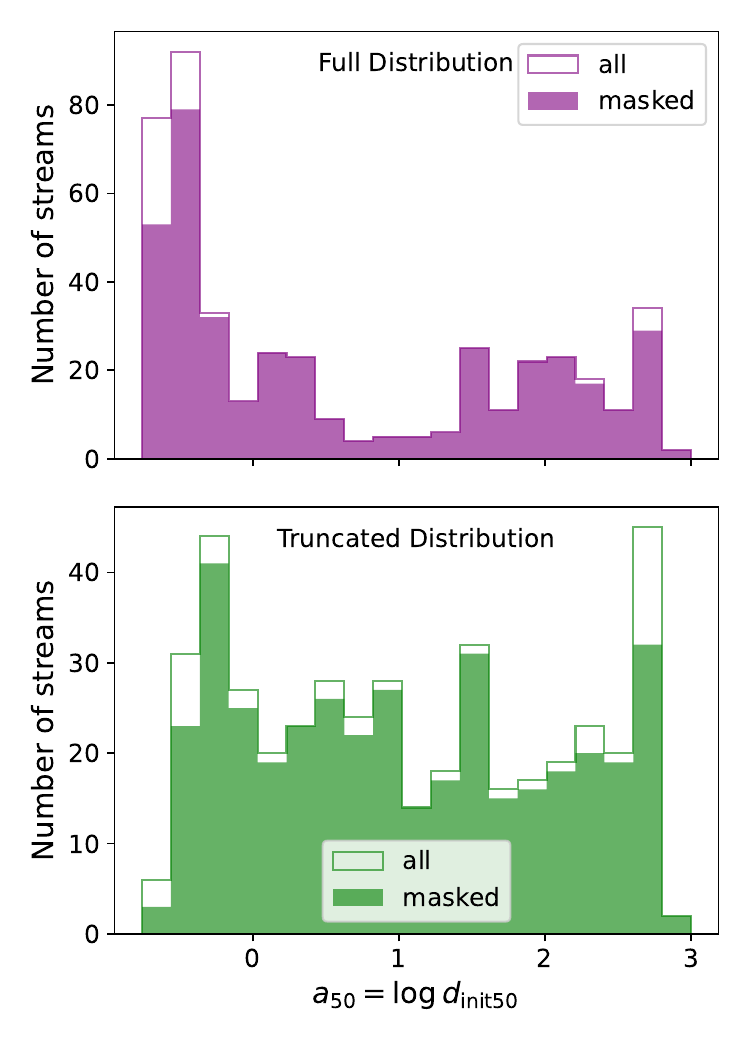}
    \caption{Same as \autoref{fig:d_init_dist} but now showing the 50th percentile of the inferred logarithmic initial sizes ($a_{50}$) after removing possible outliers from each stream’s membership list. Outliers were identified and excluded using a $3\sigma$ cut in radial velocity.}
    \label{fig:dinit50_outlier}
\end{figure}

\edit{In \autoref{subsec:obs_data}, we mention that the membership determination in \citet{hunt2024} does not incorporate radial velocities, so a small fraction of stars may be misclassified as members due to discrepant velocities. To assess the robustness of our results to this potential contamination, we repeat the analysis after removing stars whose radial velocities deviate by more than $3\sigma$ from the mean radial velocity of their associated group. We then recompute the 50th percentile of the inferred logarithmic initial sizes ($a_{50}$) for each stream. The results are shown in \autoref{fig:dinit50_outlier}. We find that excluding these potential outliers does not significantly alter the shape of the distribution of inferred initial sizes, demonstrating that our conclusions are robust to membership uncertainties in the catalogue.}

\bsp	
\label{lastpage}
\end{document}